\documentclass[12pt]{article}
\usepackage{a4wide,latexsym,graphicx,epsfig,psfrag,here}
\usepackage{amsmath,bm}
\usepackage{xcolor}
\usepackage[utf8]{inputenc}
\usepackage{amsfonts}
\usepackage{amssymb}
\usepackage[numbers,sort&compress]{natbib}
\usepackage{tikz}
\usetikzlibrary{patterns}
\usetikzlibrary{decorations}
\usepackage{comment}

\newcommand{\epsp}{\epsilon^\prime}

\newcommand{\bea}{\begin{eqnarray}}
\newcommand{\eea}{\end{eqnarray}}
\newcommand{\beq}{\begin{equation}}
\newcommand{\eeq}{\end{equation}}
\newcommand{\nn}{\nonumber}
\newcommand{\nl}{\nonumber\\}
\newcommand{\fr}{\frac}

\newcommand{\hl}{\hline}

\newcommand{\real}{{\rm Re}}
\newcommand{\imag}{{\rm Im}}

\newcommand{\cO}{{\cal O}}
\newcommand{\cL}{{\cal L}}
\newcommand{\cQ}{{\cal Q}}
\newcommand{\cA}{{\cal A}}
\newcommand{\cM}{{\cal M}}

\newcommand{\lsim}{~{}_{\textstyle\sim}^{\textstyle <}~}
\newcommand{\ba}{\begin{array}{c}}
\newcommand{\bat}{\begin{array}{cc}}
\newcommand{\ea}{\end{array}}
\def\eqn#1{(\ref{#1})}

\newcommand{\IM}{\mbox{\rm Im}}

\def\slashchar#1{\setbox0=\hbox{$#1$}\dimen0=\wd0%
\setbox1=\hbox{/}\dimen1=\wd1%
\ifdim\dimen0>\dimen1%
\rlap{\hbox to
\dimen0{\hfil/\hfil}}#1\else                                     
\rlap{\hbox to \dimen1{\hfil$#1$\hfil}}/\fi}

\begin{document}

\parskip=3pt plus 1pt

\begin{titlepage}
\mbox{}\hfill{LU TP/19-51}\\
\mbox{}\hfill{IFIC/19-46}\\
\mbox{}\hfill{DO-TH/19-23}\\
\vskip 2cm

\begin{center} 
{\LARGE \bf 
Isospin-Violating Contributions to $\boldsymbol{\epsilon'/\epsilon}$
}
\\[40pt] 
V. Cirigliano,${}^{1)}$ H. Gisbert,${}^{2)}$ A. Pich${}^{3)}$ and A. Rodr\'iguez-S\'anchez${}^{4)}$
 
\vspace{1cm}
${}^{1)}$ Theoretical Division, Los Alamos National Laboratory, \\Los Alamos, NM 87545, USA\\[10pt]

${}^{2)}$ Fakult\"at Physik, TU Dortmund,\\ Otto-Hahn-Str.4, D-44221 Dortmund, Germany \\[10pt]

${}^{3)}$ Departament de F\'{\i}sica Te\`orica, IFIC, CSIC --- 
Universitat de Val\`encia \\ 
Edifici d'Instituts de Paterna, Apt. Correus 22085, E-46071 
Val\`encia, Spain \\[10pt]

${}^{4)}$  Department of Astronomy and Theoretical Physics, \\Lund University, S\"{o}lvegatan 14A, SE 223-62 Lund, Sweden   \\[10pt]  

\end{center} 
 
\vfill 

\begin{abstract}
The known isospin-breaking contributions to the $K\rightarrow \pi\pi$ amplitudes are reanalyzed, taking into account our current understanding of the quark masses and the relevant non-perturbative inputs. We present a complete numerical reappraisal of the direct CP-violating ratio $\epsilon'/\epsilon$, where these corrections play a quite significant role. We obtain the Standard Model prediction
$\text{Re}\left(\epsilon'/\epsilon\right)\, =\,\left(14\,\pm\,5\right)\cdot 10^{-4}$,
which is in very good agreement with the measured ratio. The uncertainty, which has been estimated 
conservatively, is dominated by our current ignorance about $1/N_C$-suppressed contributions to some relevant chiral-perturbation-theory low-energy constants.

\end{abstract}

\vfill
 
\end{titlepage} 
\newpage

\tableofcontents
\newpage

\section{Introduction}

The $K\rightarrow \pi\pi$ process involves a delicate interplay between the electroweak and strong forces \cite{Cirigliano:2011ny}. At short distances the decay occurs through $W$ exchange, giving rise to a low-energy interaction between two charged weak currents. The subtleties of the strong dynamics are, however, key for understanding the decay amplitudes, even at the qualitative level, since gluonic interactions are responsible for the empirical $\Delta I=1/2$ rule that governs the measured non-leptonic decay rates, {\it i.e.}, a huge enhancement of the isoscalar $K\rightarrow \pi\pi$ amplitude over the isotensor one, 16 times larger than the naive expectation without QCD. 
Effective Field Theory (EFT) provides a powerful tool to analyze this complex dynamics, where widely separated energy scales ($M_\pi < M_K < m_c \ll M_W$) become relevant. In particular, Chiral Perturbation Theory ($\chi$PT), the EFT of the strong interactions in the low-energy regime, is ideally suited to describe $K$ decays. This work, which presents an updated study with respect to Ref. \cite{Cirigliano:2003gt}, uses this powerful EFT as theoretical framework.

While isospin symmetry is an excellent approximation for most phenomenological applications, the isospin violations induced by the quark mass difference $m_{u}-m_{d}$ and the electromagnetic interaction can get strongly enhanced in some observables \cite{Ecker:2000zr,Cirigliano:2003gt}, owing to the $\Delta I =1/2$ rule, when a tiny isospin-violating correction to the dominant amplitude feeds into the suppressed one. This is certainly the case in the direct CP-violating ratio $\epsilon'/\epsilon$, where a subtle numerical cancellation between the two isospin contributions takes place \cite{Gisbert:2017vvj}.
The current theoretical efforts to predict this observable with a precision similar to the 
experimental one \cite{Gisbert:2017vvj,Bai:2015nea,Blum:2015ywa} require an improved understanding of isospin-breaking effects \cite{Ecker:2000zr,Cirigliano:2003gt,Cirigliano:2003nn,Cirigliano:2009rr}.\footnote{For early work on this topic see Refs.~\cite{Donoghue:1986nm,Buras:1987wc,Cheng:1987dk,Lusignoli:1988fz,Wolfe:2000rf}.} This  would allow one  to test many possible New Physics (NP) scenarios that have been recently advocated \cite{Buras:2014sba,Buras:2015yca,Blanke:2015wba,Buras:2015kwd,Buras:2016dxz,Buras:2015jaq,Kitahara:2016otd,Kitahara:2016nld,Endo:2016aws,Endo:2016tnu,Cirigliano:2016yhc,Alioli:2017ces,Bobeth:2016llm,Bobeth:2017xry,Crivellin:2017gks,Chobanova:2017rkj,Bobeth:2017ecx,Endo:2017ums,Chen:2018dfc,Aebischer:2018quc,Aebischer:2018rrz,Haba:2018rzf,Matsuzaki:2018jui,Chen:2018vog,Chen:2018ytc,Haba:2018byj}. 
Re-assessing the role of the different isospin-breaking corrections is one of the main motivations of this work.

Using an isospin decomposition, the $K\to\pi\pi$ decay amplitudes can be written as\footnote{Including electromagnetic corrections, this parametrization holds for the infrared-finite amplitudes after the Coulomb and infrared parts are removed and treated in combination with real photon emission~\cite{Cirigliano:2003gt}.} \cite{Cirigliano:2003gt}
\begin{eqnarray}  
A(K^0 \to \pi^+ \pi^-) &=&  
\cA_{1/2} + {1 \over \sqrt{2}} \left( \cA_{3/2} + \cA_{5/2} \right) 
\; =\; 
 A_{0}\,  e^{i \chi_0}  + { 1 \over \sqrt{2}}\,    A_{2}\,  e^{i\chi_2 } \, ,
\nonumber\\[2pt]
A(K^0 \to \pi^0 \pi^0) &=& 
\cA_{1/2} - \sqrt{2} \left( \cA_{3/2} + \cA_{5/2}  \right) 
\; =\;
A_{0}\,  e^{i \chi_0}  - \sqrt{2}\,    A_{2}\,  e^{i\chi_2 }\, ,\label{eq:isodecomp}
\\[2pt]
A(K^+ \to \pi^+ \pi^0) &=&
{3 \over 2}  \left( \cA_{3/2} - {2 \over 3} \cA_{5/2} \right) 
\; =\;
{3 \over 2}\, A_{2}^{+}\,   e^{i\chi_2^{+}},\nonumber
\end{eqnarray}
where the three complex quantities $\cA_{\Delta I}$ are generated by the $\Delta I = 1/2,\:3/2,\:5/2$ components of the electroweak effective Hamiltonian, in the limit of isospin conservation. In that limit, $A_0$
and $A_2=A_2^+$ denote the decay amplitudes into $(\pi\pi)_I$ states with $I=0$ and $2$, while the phases $\chi_0$ and $\chi_2=\chi_2^+$ are the S-wave 
$\pi \pi$ scattering phase shifts at $\sqrt{s} = M_K$. By definition, the amplitudes $A_I$ are real and positive in the CP-conserving limit.  From the measured $K\to\pi\pi$ branching ratios, one finds \cite{Antonelli:2010yf}
\begin{eqnarray}
A_0 &=& (2.704 \pm 0.001) \cdot 10^{-7} \mbox{ GeV},\nonumber \\
A_2 &=& (1.210 \pm 0.002) \cdot 10^{-8} \mbox{ GeV}, \label{eq:isoamps}\\
\chi_0 - \chi_2 &=& (47.5 \pm 0.9)^{\circ}.\nonumber
\end{eqnarray}

When CP violation is turned on, the amplitudes $A_0$, $A_2$ and $A_2^+$ acquire imaginary parts and $\epsilon'$ is given to first order in CP violation by 
\begin{equation}
\epsilon'\,  =\, 
- \frac{i}{\sqrt{2}} \: e^{i ( \chi_2 - \chi_0 )} \:\omega\;
\left[
\frac{\mathrm{Im} A_{0}}{ \mathrm{Re} A_{0}} \, - \,
\frac{\mathrm{Im} A_{2}}{ \mathrm{Re} A_{2}} \right] 
\, =\, 
- \frac{i}{\sqrt{2}} \: e^{i ( \chi_2 - \chi_0 )} \:\omega\;
\frac{\mathrm{Im} A_{0}}{ \mathrm{Re} A_{0}} \,\bigg( 1\:-\:\frac{1}{\omega}\;\frac{\mathrm{Im} A_{2}}{\mathrm{Im} A_{0}}\bigg) \, .
\label{eq:eps1}
\end{equation}
Then, $\epsilon'$ is suppressed by the ratio $\omega \equiv\mathrm{Re} A_2 / \mathrm{Re} A_0\approx 1/22$ and $\epsilon'/\epsilon$ is approximately real, since 
$\chi_2-\chi_0-\phi_{\epsilon}\approx 0$,  being $\phi_{\epsilon}$ the superweak phase. Moreover, the last expression makes manifest the important potential role of isospin-breaking effects. Any small correction to the ratio $\mathrm{Im} A_{2}/\mathrm{Im} A_{0}$ gets amplified by the large value of $\omega^{-1}$.

It is well known that the further chiral enhancement of the electromagnetic penguin contributions to $\mathrm{Im} A_{2}$ makes compulsory taking them into account for any reliable estimate of $\epsilon'/\epsilon$, in spite of the fact that they are isospin-violating corrections. Futhermore, 
Eq.~\eqref{eq:eps1} contains a delicate numerical balance between the two  isospin contributions,  making the result very sensitive to any additional isospin-breaking corrections. 
Indeed, simplified estimates of $\mathrm{Im} A_{I}$ result in a strong cancellation between  the  two  terms,  leading  to very  low values for $\epsilon'/\epsilon$ \cite{Buras:1993dy,Buras:1996dq,Bosch:1999wr,Buras:2000qz,Ciuchini:1995cd,Ciuchini:1992tj,Buras:2015xba,Buras:2016fys,Buras:2015yba}.
A critique  of these approaches has been recently presented in  Ref.~\cite{Gisbert:2017vvj}.
A proper assessment of the isospin-violating contributions to the $K\to\pi \pi$ amplitudes is then a compulsory requirement for making reliable predictions of $\epsilon'/\epsilon$.

A detailed study of isospin-breaking effects in $K\to\pi \pi$ was performed in Ref.~\cite{Cirigliano:2003gt,Cirigliano:2003nn,Cirigliano:2009rr}. 
While the analytical calculations reported in these references remain valid nowadays, meanwhile
there have been many relevant improvements in the needed inputs that make worth to perform an updated analysis of their phenomenological implications.  The much better precision achieved in the determination of quark masses allows now for improved estimates of the penguin matrix elements. Moreover, we have at present a better understanding of several non-perturbative ingredients such as the chiral Low-Energy Constants (LECs), which govern the $\chi$PT $K\to\pi\pi$ amplitudes
\cite{Aoki:2016frl,Ecker:1988te,Ecker:1989yg,Pich:2002xy,Cirigliano:2006hb,Kaiser:2007zz,Cirigliano:2004ue,Cirigliano:2005xn,RuizFemenia:2003hm,Jamin:2004re,Rosell:2004mn,Rosell:2006dt,Pich:2008jm,GonzalezAlonso:2008rf,Pich:2010sm,Bijnens:2014lea,Rodriguez-Sanchez:2016jvw,Ananthanarayan:2017qmx}. Implementing those improvements by updating Ref.~\cite{Cirigliano:2003gt} is one of the main motivation for this work.

In Section \ref{sec:EFTdesc}, we review the different low-energy Lagrangians involved in the $K\rightarrow \pi\pi$ process. We describe the structure of the amplitudes at next-to-leading order (NLO) in $\chi \text{PT}$, including
isospin-breaking corrections, in Section~\ref{sec:ampnlo}. The main limitation of the $\chi$PT approach originates in the not very well-known LECs that encode  
dynamical information from the non-perturbative QCD scale $\sim 1$~GeV.
 Our current knowledge on those LECs is compiled in Section~\ref{sec:detLEC}. Section~\ref{sec:Decomp} gives the chiral expansion of the different isospin amplitudes to first order in isospin-breaking and CP violation.
Finally, we present the numerical results  in Section~\ref{sec:numbers} and 
discuss their impact on $\epsilon^\prime/\epsilon$ in Section~\ref{sect:conclusions}.
We provide some technical details in a set of appendices.

\section{Effective field theory description}\label{sec:EFTdesc}

At the electroweak scale, the $\Delta S = 1$ transition is described in terms of quarks and gauge bosons. Owing to the different mass scales involved, 
the gluonic corrections are amplified with large logarithms, such as $\log(M_W/m_c)\sim 4$, 
that can be summed up all the way down to scales $\mu_{\rm SD}<m_c$, using the Operator Product Expansion (OPE) and the Renormalization Group Equations (RGEs). One obtains in this way a short-distance effective $\Delta S = 1$ Lagrangian, defined in the three-flavour theory \cite{Buchalla:1995vs},
\begin{align}
\cL_{\rm eff}^{\Delta S = 1}\:=\:-\frac{G_F}{\sqrt{2}}\:V_{ud}\:V_{us}^*\sum_{i=1}^{10} C_i(\mu_{\rm SD})\:Q_i(\mu_{\rm SD})~,\label{eq:shortLagr}
\end{align}
which is a sum of local four-quark operators $Q_i$,  weighted by Wilson coefficients $C_i(\mu_{\rm SD})$. that are functions of the heavy masses ($M_Z, M_W, m_t, m_b, m_c$) and CKM parameters:
\begin{equation}
 C_i(\mu_{\rm SD})\, =\, z_i(\mu_{\rm SD}) +\tau\: y_i(\mu_{\rm SD}) \, , 
 \qquad\qquad
\tau\, =\, -\frac{V_{td}V_{ts}^*}{V_{ud}V_{us}^*} \, .
\end{equation}

The CP-violating effects originate in the CKM ratio $\tau$ and are thus governed by the $y_i(\mu_{\rm SD})$ short-distance coefficients, while the $K\to\pi\pi$ amplitudes are fully dominated by the CP-conserving factors $z_i(\mu_{\rm SD})$. 
These Wilson coefficients are known to NLO \cite{Buras:1991jm,Buras:1992tc,Buras:1992zv,Ciuchini:1993vr}, which includes all corrections of $\cO(\alpha_s^n t^n)$ and 
$\cO(\alpha_s^{n+1} t^n)$ with $t\equiv\log{(M_1/M_2)}$ the logarithm of any ratio of heavy mass scales. The complete calculation of next-to-next-to-leading (NNLO) QCD corrections is expected to be finished soon \cite{Cerda-Sevilla:2016yzo,Buras:1999st,Gorbahn:2004my}.

The renormalization scale ($\mu_{\rm SD}$) and scheme dependence of the $C_i(\mu_{\rm SD})$ coefficients should exactly cancel with a corresponding dependence of the hadronic matrix elements $\langle \pi\pi | Q_i(\mu_{\rm SD}) | K\rangle$. Unfortunately, a rigorous analytic evaluation of these non-perturbative matrix elements, keeping full control of the QCD renormalization conventions, remains still a very challenging task. Nevertheless, we can take advantage of the symmetry properties of the four-quark operators to build their low-energy realization within the $\chi$PT framework. 
The difference $Q_- \equiv Q_2-Q_1$ and the QCD penguin operators $Q_{3,4,5,6}$ induce pure $\Delta I=\frac{1}{2}$ transitions and transform as $(8_L,1_R)$ under chiral $SU(3)_L\otimes SU(3)_R$ flavour transformations. Transition amplitudes with $\Delta I=\frac{3}{2}$ can only be generated by the complementary combination $Q^{(27)}\equiv 2 \, Q_2+ 3\, Q_1 - Q_3$, which transforms as a $(27_L,1_R)$ operator and can also induce $\Delta I=\frac{1}{2}$ transitions. The electroweak penguin operators do not have definite isospin and chiral quantum numbers, due to their explicit dependence on the light-quark electric charges $e_q$. $Q_7$ and $Q_8$ can be split into combinations of $(8_L,1_R)$ and $(8_L,8_R)$ pieces, while $Q_9$ and $Q_{10}$ contain $(8_L,1_R)$ and $(27_L,1_R)$ components.

\subsection[$\chi$PT formulation]{$\boldsymbol{\chi}$PT formulation}\label{subsec:chpt}

Chiral symmetry allows one to formulate another EFT, $\chi$PT, that is valid at the kaon mass scale where perturbation theory cannot be trusted. The Goldstone nature of the lightest octet of pseudoscalar mesons strongly constrains  their interactions \cite{Pich:2018ltt}, providing a very powerful tool to describe kaon decays in a rigorous way \cite{Cirigliano:2011ny}. Knowing the symmetry properties of the relevant QCD amplitudes, one can build their effective $\chi$PT realization in terms of the pseudoscalar meson fields as systematic expansions in powers of momenta, $p^2$, quark masses, $m_{q}$, and electric charges, $e_q^{2}$. According to the Weinberg power-counting theorem \cite{Weinberg:1978kz}, loop corrections introduce extra powers of $p^{2}$, so that they enter at the same level as higher-order operators. All the short-distance information about the heavy particles that have been integrated out of the low-energy EFT is encoded in the LECs of the $\chi$PT Lagrangian.

In the following, we compile the relevant effective Lagrangians associated to the different interactions entering in our $K\to\pi\pi$ analysis. Further details about the strong Lagrangian at $\cO(p^6)$~\cite{Gasser:1984gg,Fearing:1994ga,Bijnens:1999sh}, the nonleptonic weak Lagrangian to $\cO(G_F p^4)$~\cite{Cronin:1967jq,Kambor:1989tz,Ecker:1992de,Bijnens:1998mb}, the electromagnetic Lagrangian to $\cO(e^2 p^2)$~\cite{Ecker:1988te,Urech:1994hd} and the electroweak Lagrangian to $\cO(e^2 G_8 p^2)$~\cite{Bijnens:1983ye,Grinstein:1985ut,Ecker:2000zr} can be found in the quoted references.

The strong $\chi$PT Lagrangian is given by\footnote{The $\cO(p^6)$ LECs are usually denoted $C_i\equiv F^{-2} X_i$. We have changed the notation to avoid possible confusions with the short-distance Wilson coefficients.}
\begin{equation}
\cL_{\rm strong}  \; =\; \frac{F^2}{4} \,\langle D_\mu U D^\mu U^\dagger + 
\chi U^\dagger + \chi^\dagger  U \rangle
 +  \sum_{i=1}^{10}\; L_i\, O^{p^4}_i +  F^{-2}\sum_{i=1}^{90}\; X_i\, O^{p^6}_i
 + \cO(p^8)~, 
 \label{eq:Lstrong}
\end{equation}
where $U(x)\equiv\exp{\{i \lambda^a \phi^a(x)/F\}}$ is the $\text{SU}(3)$ unitary matrix that parametrizes the pseudoscalar fields, $D_\mu U$ is the covariant derivative matrix, $\chi\equiv 2 B_0\cM$ takes into account the explicit chiral symmetry breaking through the quark mass matrix $\cM =\mathrm{diag}(m_u,m_d,m_s)$, 
and $\langle \cdots \rangle$ indicates an $\text{SU}(3)$ flavour trace.
The different pieces correspond, respectively, to $\cO(p^2)$, $\cO(p^4)$ and $\cO(p^6)$ in the chiral expansion. Notice how the number of LECs increases with the $\chi$PT order.

To $\cO(G_F p^4)$, the nonleptonic $\Delta S=1$ weak interactions are described by  
\begin{eqnarray} 
\cL^{\Delta S=1}
&=&  G_8 \, F^4  \,\langle\lambda D^\mu U^\dagger
 D_\mu U \rangle  + G_8\, F^2\sum^{22}_{i=1}   N_i \, O^8_i \nn\\
& + & G_{27}\, F^4 \left( L_{\mu 23} L^\mu_{11} + 
{2\over 3} L_{\mu 21} L^\mu_{13}\right)  +
  G_{27}\, F^2\sum^{28}_{i=1} D_i \, O^{27}_i + \cO(G_{F} p^6)\, , 
\label{eq:Lweak}
\end{eqnarray}
where $\lambda=(\lambda_6-i\, \lambda_7)/2$ projects onto the $\bar{s}\rightarrow\bar{d}$ transition and $L_\mu\:=\:i\:U^\dagger\:D_\mu\:U$ represents the octet of $\text{V}-\text{A}$ currents to lowest order in derivatives. Under chiral transformations, the first and the second lines of Eq.~(\ref{eq:Lweak}) transform as $(8_L,1_R)$ and $({27}_L,1_R)$, respectively, providing the effective low-energy realization of the $Q_{i\le 6}$ components in Eq.~(\ref{eq:shortLagr}). The first term of each line corresponds to $\cO(G_{F} p^2)$, while the second one to $\cO(G_{F} p^4)$. The explicit list of relevant operators $O^{8}_i$ and $O^{27}_i$ for $K\rightarrow \pi\pi$ can be found in the Appendix A of Ref.~\cite{Cirigliano:2003gt}. Furthermore, to simplify the notation, we introduce the dimensionless couplings $g_8$ and $g_{27}$, defined as
\begin{equation}
G_{8,27}\,\equiv\, -\frac{G_F}{\sqrt{2}}\; V_{ud}V_{us}^* \:g_{8,27}.
\end{equation}
In Eq.~(\ref{eq:Lweak}), there are 52 dimensionless LECs: $g_8$, $g_{27}$, $N_i$ and $D_i$. In Section~\ref{sec:detLEC}, we will explain how to estimate these couplings using large-$N_C$ techniques. 

The electromagnetic Lagrangian starts at $\cO(e^2p^0)$. Including $\cO(e^2p^2)$ terms, one has:
\begin{equation} 
\cL_{\rm elm} \; =\;  e^2\, Z \, F^4  \,\langle \cQ  U^\dagger \cQ U\rangle 
+ e^2 \,F^2\, \sum^{14}_{i=1}\; K _i \, O^{e^2 p^2}_i + \cO(e^2p^4)~. 
\label{eq:Lelm}
\end{equation}
where $\cQ = \text{diag}(2/3,\:-1/3,\:-1/3)$ is the quark charge matrix and $Z$ is the lowest-order LEC that is related,  up to $ {\cal O}(e^2 m_q)$ corrections, 
to the pion mass difference
\begin{align}
Z\;\approx\;\frac{1}{8\pi\alpha F^2}\; (M_{\pi^\pm}^2\:-\:M_{\pi^0}^2)\;\approx\; 0.8~.
\end{align}
The NLO LECs $K_i$ are dimensionless and explicit expressions for those operators $ O^{e^2 p^2}_i$ that are relevant in $K\rightarrow \pi\pi$ can be found in the Appendix A of Ref.~\cite{Cirigliano:2003gt}.

Finally, the relevant $\Delta S=1$ electroweak Lagrangian contains $\cO(e^2G_{F}p^0)$ and $\cO(e^2G_{F}p^2)$ terms:
\begin{equation} 
\cL^{\Delta S=1}_{\rm EW} \; =\; e^2 \,G_8 \, g_{\rm ewk}\, F^6 \,\langle\lambda U^\dagger \cQ 
U\rangle 
 +   e^2   \, G_8\, F^4\,\sum^{14}_{i=1} Z_i\, O^{EW}_i + \cO(G_F e^2 p^4)\, .
\label{eq:Lelweak} 
\end{equation}
This Lagrangian transforms as $(8_L,8_R)$ under chiral transformations and provides the needed low-energy realization of the electromagnetic penguin operators in Eq.~(\ref{eq:shortLagr}). Notice that we will not include isospin-violating corrections for the 27-plet amplitudes and, therefore, the electroweak $(27_L,1_R)$ chiral structures are not needed. The LECs $Z_i$ are dimensionless and the associated operators $O^{EW}_i$ are collected in Appendix A of Ref.~\cite{Cirigliano:2003gt}.

At the chiral order we are working in, all loop divergences are reabsorbed by the previous LECs ($\mathcal{C}_i=L_i,\:N_i,\:D_i,\:K_i,\:Z_i$), which have to be renormalized. At one-loop, they can be expressed as
\begin{align}
\mathcal{C}_i \; =\; \mathcal{C}_i^r (\nu_\chi) + c_i \,\Lambda (\nu_\chi) \, ,
\end{align}
where $\nu_\chi$ is the chiral renormalization scale and the divergence is included in the factor
\begin{align}
\Lambda (\nu_\chi)\; =\; \frac{\nu_\chi^{d-4}}{(4 \pi)^2} \, \left\{ 
\frac{1}{d-4} - \frac{1}{2} 
\bigg[ \log (4 \pi) + \Gamma ' (1) + 1 \bigg] 
\right\} \ .  
\label{eq:div}
\end{align}
The divergent parts of all these couplings ($c_i=\Gamma_i,\:n_i,\:d_i,\:\kappa_i,\:z_i$) are known and can be found in Ref.~\cite{Gasser:1984gg,Kambor:1989tz,Ecker:1992de,Urech:1994hd,Ecker:2000zr}, respectively.

\section{$\boldsymbol{K\to\pi\pi}$ amplitudes up to  NLO}
\label{sec:ampnlo}

Once the different effective chiral Lagrangians involved in $K\to\pi\pi$ have been introduced, we are in position to obtain the physical amplitudes, using the $\chi$PT power-counting rules. For the isospin conserving parts, {\it i.e.}, when $e^2 = m_{u}-m_{d}=0$, the $\cO(G_{F}p^2)$ contributions to the $\cA_{\Delta I}$ amplitudes defined in Eq.~\eqn{eq:isodecomp} are given by 
\begin{eqnarray}
\cA_{1/2} & = &  - \sqrt{2}\, G_8 F\, \Big[  \left( M_{K}^2 - M_{\pi}^2 \right)
 \Big] - {\sqrt{2} \over 9}\, G_{27} F \left( M_{K}^2 - M_{\pi}^2 \right)  ,\nonumber
\\
\cA_{3/2} & = & -
{10 \over 9}  G_{27} F \left( M_{K}^2 - M_{\pi}^2 \right),\label{eq:LOamp}\\
\cA_{5/2} & = & 0\, .\nonumber
\end{eqnarray}
Using the measured amplitudes in Eq.~\eqn{eq:isoamps}, one immediately obtains the tree-level determinations $g_8 = 5.0$ and $g_{27} = 0.25$ for the octet and 27-plet chiral couplings, respectively. The large numerical difference between these two LECs reflects the smallness of the measured ratio 
\begin{eqnarray}
\omega  = \frac{\cA_{3/2}}{\cA_{1/2}} \approx \frac{1}{22}\, ,
\label{eq:omegaw}
\end{eqnarray}
known as the $\Delta I=1/2$ rule. 

In this work, we use the full  $\cO(G_{F}p^4)$ expressions for the isospin-conserving parts of the amplitudes. Isospin-breaking corrections are accounted only at first order, {\it i.e.}, only corrections of $\cO(e^2(m_{d}-m_{u})^0)$ and $\cO(e^0(m_{d}-m_{u}))$ are considered. Additionally, owing to the very small value of $g_{27}/g_8$,  and the fact that  $\text{Im}(g_{27})=0$ in the large-$N_C$ limit, we neglect isospin-breaking corrections proportional to $g_{27}$, which have been calculated in \cite{Bijnens:2004ai}.
We outline below the relevant sources of isospin breaking up to NLO in  $\chi$PT.

\subsection{Leading Order}

To lowest order in the number of derivatives and quark mass insertions 
the sources of isospin breaking are (i)  the term in $\cL_{\rm strong}$ with one quark mass insertion; 
(ii)  the non-derivative term in $\cL_{\rm elm}$, proportional to $e^2Z$; and
(iii) the non-derivative term in $\cL^{\Delta S=1}_{\rm EW}$, proportional to $e^2 G_8 \, g_{\rm ewk}$.
Sources (i) and (ii) affect the pseudoscalar meson mass matrix generating two effects:  

\begin{itemize}
\item  $\pi^0-\eta$ mixing, due to non-diagonal terms coupling the SU(3) fields $\pi_3$ and $\eta_8$: 
\beq 
\left( \begin{array}{c} \pi_3 \\ \eta_8  \end{array} \right)\; =\; 
\left( \begin{array}{cc} 1 &  - \varepsilon^{(2)} \\ 
\varepsilon^{(2)}  & 1  \end{array} \right) \, 
\left( \begin{array}{c} \pi^0 \\ \eta  \end{array} \right)
\, .  
\label{eq:LOmixing}
\eeq
The  tree-level mixing angle is given by 
\beq
\varepsilon^{(2)}\; =\; \frac{\sqrt{3}}{4} \; 
\frac{m_d - m_u}{m_s - \widehat{m}}\,\equiv\,\frac{\sqrt{3}}{4 \,R} \;  =\; (1.137\pm 0.045)\cdot 10^{-2} \, ,
\label{epsilon}
\eeq 
where $\widehat{m} = (m_u+m_d)/2$. 
We have extracted the numerical value from the most recent FLAG average of lattice determinations of light-quark masses, with $N_f=2+1$ dynamical fermions, which quotes
$R = 38.1\pm 1.5$ \cite{Aoki:2019cca}.

\item Mass splitting   between charged and neutral mesons, due to 
both the light quark mass difference and electromagnetic contributions.  
Following Ref~\cite{Cirigliano:2003gt}, we choose to express all masses in terms of those of the neutral kaon and pion (denoted from now on as $M_K$ and $M_\pi$, respectively).
In terms of quark masses and LO couplings ($B_0$ is related 
to the quark condensate in the chiral limit by $\langle 0 | 
\overline{q} q | 0 \rangle = -F^2 B_0$), 
up to corrections of $\cO(m_q^2, e^2 m_q)$ 
the pseudoscalar meson masses 
read:
\bea
 M^2_{\pi} &=& 2 B_0 \,\widehat{m}\, ,   \nl
 M^2_{\pi^\pm} &=& M_\pi^2 + 2\, e^2 Z F^2\, ,   \nl
 M^2_{K} &=& B_0 \left( m_s + m_d \right)\, ,  \label{eq:treemass} \\
 M^2_{K^\pm} &=& M_K^2 
-  \frac{4 \,\varepsilon^{(2)}}{\sqrt{3}}\,  B_0 (m_s - \widehat{m})  
+ 2\, e^2 Z F^2\, ,    \nl
 M^2_\eta &=& \frac{1}{3} \, \left( 4 M_K^2  - M_\pi^2 \right) 
-  \frac{8 \,\varepsilon^{(2)}}{3 \sqrt{3}}\,  B_0 (m_s - \widehat{m})
~.\nn
\eea
The above choice defines a specific  ``isospin limit  scheme'', 
which is however arbitrary.  
In Appendix \ref{app:scheme} we explore another  quite natural 
scheme and quantify the impact of such scheme dependence on $\epsilon^\prime / \epsilon$. 
We find that the scheme dependence is well below the current theoretical uncertainties. 
\end{itemize}

The sources of isospin breaking described above induce corrections to 
the  $K \rightarrow \pi \pi$  amplitudes of  $\cO(\varepsilon^{(2)}\, G_{8}\, p^{2})$ and  $\cO(e^{2}\, G_{8}\, p^{0})$.
Explicitly,  the three independent  $K \rightarrow \pi \pi$  
amplitudes  in the isospin basis read: 
\bea 
\cA_{1/2} & = & 
- {\sqrt{2} \over 9}\, G_{27} F \left( M_K^2 - M_\pi^2 \right)  \nn \\
 & &  -
\sqrt{2}\, G_8 F \left[  \left( M_K^2 - M_\pi^2 \right) 
 \left(1 - {2 \over 3 \sqrt{3}} \,\varepsilon^{(2)} \right)  
- {2 \over 3}\, e^2 F^2 \left( g_{\rm ewk} + 2 Z \right)  \right]\, ,
\nonumber 
\\
\cA_{3/2} & = & 
- {10 \over 9}\,  G_{27} F \left( M_K^2 - M_\pi^2 \right) - 
G_8 F \left[  \left( M_K^2 - M_\pi^2 \right) {4 \over 3  \sqrt{3}}\, 
\varepsilon^{(2)}  -  {2 \over 3}\, e^2 F^2 \left( g_{\rm ewk} + 2 Z \right)
\right]\, , \nn   \\ 
\cA_{5/2} & = & 0 ~. 
\label{eq:LOamp2} 
\eea
The parameter $F$ can be identified with the pion
decay constant $F_\pi$ at this order.  The effect
of strong isospin breaking (proportional to $\varepsilon^{(2)}$) is
entirely due to $\pi^0 - \eta$ mixing,  through expressing all  interaction vertices in terms of mass eigenfields. 
Electromagnetic interactions contribute through mass splitting in the external legs (terms 
proportional to $Z$) and insertions of $g_{\rm ewk}$.

\subsection{Next-to-Leading Order}

NLO isospin-breaking corrections due to loops and effective Lagrangians with  additional powers of derivatives and quark mass insertions  ($\cO (\varepsilon^{(2)} G_8 p^4, e^2 G_8 p^2)$)   
generate many new contributions: 
\begin{itemize}
\item $\cO(\varepsilon^{(2)} \, G_{8} \, p^4)$. One has:
\begin{itemize}
\item$\pi^0-\eta$ mixing at NLO. Identical to the previous correction but changing $\varepsilon^{(2)}\rightarrow\varepsilon^{(4)}_S$ \cite{Ecker:1999kr,Cirigliano:2003gt},
\begin{eqnarray}
\varepsilon^{(4)}_{\rm S} &= &
- \frac{2 \, \varepsilon^{(2)}}{3 (4 \pi F)^2 (M_{\eta}^2 - 
M_{\pi}^2)} 
  \bigg\{ (4 \pi)^2 \, 64 \left[3 L_7 + 
L_8^r (\nu_\chi) \right] 
(M_K^2 - M_\pi^2)^2 
\nl
&& {} -  M_\eta^2 (M_K^2 - M_\pi^2) \log \frac{M_\eta^2}{\nu_\chi^2}
 +  M_\pi^2 (M_K^2 - 3 M_\pi^2) \log \frac{M_\pi^2}{\nu_\chi^2}  \nn \\
&& {} - 2 M_K^2 (M_K^2 - 2 M_\pi^2) \log \frac{M_K^2}{\nu_\chi^2} 
- 2 M_K^2 (M_K^2 - M_\pi^2) \bigg\}~.
\label{eq:epsilon4S}
\end{eqnarray}

\item  Diagrams with isospin-conserving vertices and isospin-breaking corrections to the pseudoscalar masses, either in the propagators or the on-shell external legs.

\item Diagrams analogous to the isospin-conserving ones, but with vertices obtained after applying the rotation of Eq. (\ref{eq:LOmixing}), so that one of the vertices introduces an $\varepsilon^{(2)}$ factor.

\end{itemize}

\item $\cO(e^2\, G_{8}\, p^2)$, entering through:

\begin{itemize}
\item$\pi^0-\eta$ mixing at NLO. Identical to the strong isospin-breaking correction but with $\varepsilon^{(2)}\rightarrow \varepsilon_{EM}^{(4)}$ \cite{Cirigliano:2001mk,Cirigliano:2003gt},
\begin{eqnarray}
\varepsilon^{(4)}_{\rm EM} &=&   
\frac{2 \, \sqrt{3} \, \alpha }{108 \, \pi \, (M_\eta^2 
-M_\pi^2)}  
  \bigg\{ 
-  9 M_K^2   Z \left(\log \frac{M_K^2}{\nu_\chi^2} + 1 \right) \nl
&& {} + 2  M_K^2  (4 \pi)^2 \Big[ 2 U_{2}^{r} (\nu_\chi) + 
3 U_{3}^{r} (\nu_\chi) \Big] \nl
&& {} 
+  M_\pi^2  (4 \pi)^2 \Big[ 2 U_{2}^{r} (\nu_\chi) + 
3 U_{3}^{r} (\nu_\chi) - 6 U_{4}^{r} (\nu_\chi) \Big]  \bigg\}
 ~,
\end{eqnarray}
where $U_{i}^{r} (\nu_\chi)$ are linear combinations of the $K_{i}^{r}$ LECs defined in Eq.~(\ref{eq:Lelm}),
\begin{align}
U_1&=K_1+K_2\, ,\qquad\qquad\:\:\: U_2=K_5+K_6\, ,
\nonumber\\[0.25cm]    
U_3&=K_4-2\: K_3\, ,\qquad\qquad U_4=K_9+K_{10}\, .  
\end{align}
\item Loop corrections with one $g_{8}\, g_{\rm ewk}$ vertex.
\item Again, diagrams with isospin-conserving vertices and  isospin-breaking corrections to the pseudoscalar masses either in the propagators or the external legs.
\item Electromagnetic loop corrections with one $g_{8}$ vertex and virtual photon propagators. In order to cancel the infrared divergences, one must also add the corresponding calculation of the $K\to\pi\pi\gamma$ rates \cite{Cirigliano:2003gt}.
\item Tree-level diagrams with at least one electroweak vertex and a NLO insertion.
\end{itemize}

\end{itemize}

\subsection{Structure of the   amplitudes up to NLO}
Taking into account the previous discussion, the isospin amplitudes $\cA_n$ ($n=1/2,\:3/2,\:5/2$) can be expressed  as
\begin{align}
\cA_n &= - G_{27} \, F_\pi \, \Big( M_K^2 - M_\pi^2 \Big) \,  \cA_{n}^{(27)} - 
G_{8} \, F_\pi \,
\Big( M_K^2 - M_\pi^2 \Big) 
\Big[ \cA_{n}^{(8)} +  \varepsilon^{(2)} \, \cA_{n}^{(\varepsilon)} \Big] \nn\\
& +   e^2 \,G_{8} \, F_\pi^3 \,  \Big[ \cA_{n}^{(\gamma)} + Z \,  \cA_{n}^{(Z)} + 
 g_{\rm ewk} \,  \cA_{n}^{(g)} \Big] ~,\label{eq:generalamplitude}
\end{align}
where $\cA_{n}^{(\varepsilon)}$ refers to the strong isospin breaking contributions, $\cA_{n}^{(g)}$ and $\cA_{n}^{(Z)}$ are the contributions with an insertion of  $g_{\rm ewk}$  and $Z$ vertices, and $\cA_{n}^{(\gamma)}$ are the contributions induced by the photon loops. In Eq.~(\ref{eq:generalamplitude}), we have replaced the Goldstone coupling $F$ by $F_\pi$, the physical pion decay constant at NLO. These two parameters are related through \cite{Gasser:1984gg,Knecht:1999ag}
\begin{align}\label{eq:Fpi_F}
F &= F_\pi \;\Bigg\{ 1 - \frac{4}{F^2} \Bigg[ L_{4}^{r} (\nu_\chi)  
\left( M_\pi^2 + 2 M_K^2 \right) + L_{5}^{r}(\nu_\chi)  M_\pi^2 \Bigg] 
\nn \\ 
&\hskip 1.1cm + 
\frac{1}{2 (4 \pi)^2 F^2} \left[ 2 \,M_\pi^2 \log\bigg( \frac{M_\pi^2}
{\nu_\chi^2}\bigg) + 
M_K^2 \log \bigg(\frac{M_K^2}{\nu_\chi^2}\bigg) \right] 
\nn \\
&\hskip 1.1cm +  
   \frac{2\, \varepsilon^{(2)}}{\sqrt{3}} \left( M_K^2 - M_\pi^2 
\right)  \, \left[ \frac{8 L_4^r (\nu_\chi)  }{F^2} - 
\frac{1}{2 (4 \pi)^2 F^2} 
\left( 1 +  \log\bigg( \frac{M_K^2}{\nu_\chi^2}\bigg) \right)  \right] 
\Bigg\} \ ,
\end{align}
so that those corrections get reabsorbed into the different NLO terms. 

Each  amplitude $\cA_{n}^{(X)}$ in Eq.~(\ref{eq:generalamplitude}) can be decomposed as
\begin{align}
\cA_{n}^{(X)}\; =\; \left\{ \begin{array}{cll}
a_n^{(X)} \, \left[ 1  + \Delta_{L} \cA_{n}^{(X)} + \Delta_{C} \cA_{n}^{(X)}
\right] ,  & \qquad  \mbox{if} \   &  a_n^{(X)} \neq 0\, ,
\label{eq:structure2} \\
\Delta_{L} \cA_{n}^{(X)} + \Delta_{C} \cA_{n}^{(X)}\, , &
\qquad \mbox{if} \  & 
 a_n^{(X)} = 0 \, ,
\end{array}
\right.
\end{align}
with $a_n^{(X)}$, $\Delta_{L} \cA_{n}^{(X)}$ and $\Delta_{C} \cA_{n}^{(X)}$ being the LO, NLO loop and NLO local contributions, respectively.\footnote{
Strictly speaking, by  expressing the tree-level amplitudes in terms of physical meson masses and  $F_\pi$, the term dubbed as ``LO" contains NLO chiral corrections. 
While the splitting of LO and NLO terms is indeed ambiguous, our amplitudes are correct up to and including terms of order $G_F p^4$, $G_F \varepsilon^{(2)} p^4$, 
and $G_F e^2 p^2$.}
The amplitudes $\cA_{n}^{(X)}$ and their components $a_n^{(X)}$, $\Delta_{L} \cA_{n}^{(X)}$ and $\Delta_{C} \cA_{n}^{(X)}$ are dimensionless by construction.
In Table~\ref{tab:LOcontributions}, we give the values of the LO factors $a_n^{(X)}$. 
The loop corrections $\Delta_{L} \cA_{n}^{(X)}$ account for the requirements of unitarity and analyticity; these non-local contributions are fully predicted in terms of the pseudoscalar masses and the pion decay constant. The local components $\Delta_{C} \cA_{n}^{(X)}$ contain the explicit dependence on the NLO LECs that renormalize the ultraviolet loop divergences. Therefore, both $\Delta_{L} \cA_{n}^{(X)}$ and $\Delta_{C} \cA_{n}^{(X)}$ depend on the $\chi$PT renormalization scale, but this dependence exactly cancels in their sum.
The full expressions for $\Delta_{L} \cA_{n}^{(X)}$ and $\Delta_{C} \cA_{n}^{(X)}$ can be found in Appendix B and in Section 4.4 of Ref.~\cite{Cirigliano:2003gt} respectively. 

\begin{table}[tb] 
\begin{center}
\begin{tabular}{|c|c|c|c|c|c|c|}\hl 
n & (27) &
(8) &
($\varepsilon$)   & ($Z$) & ($g$) \\
\hl
1/2  & 
${\sqrt{2} \over 9}$   & 
$\sqrt{2}$   & 
$-\frac{2}{3}\frac{\sqrt{2}}{\sqrt{3}}$  & $\frac{4 \sqrt{2}}{3}$ & $\frac{2 \sqrt{2}}{3}$  \\
\hl
3/2  &  
${10 \over 9}$   & 
$0$   & 
$\frac{4}{3\sqrt{3}}$  & $\frac{4 }{3}$ & $\frac{2 }{3}$
\\
\hl
\end{tabular}
\end{center} 
\caption{LO contributions $a_n^{(X)}$ for $n=1/2,\:3/2$. $a_{5/2}^{(X)}=0$ for all $X$ and $a_{n}^{(\gamma)}=0$ for all $n$.}  
\label{tab:LOcontributions}
\end{table}

\section{Determination of chiral LECs}\label{sec:detLEC}

In the last section, we have introduced the general structure of the $K\to\pi\pi$ amplitudes up to  NLO. The only remaining ingredients are the $\chi$PT LECs, which are not fixed by symmetry considerations.

\begin{figure}[h!]\centering
\setlength{\unitlength}{0.9mm} 
\resizebox{1.2\totalheight}{!}{
\begin{picture}(156,122)
\put(0,0){\makebox(156,120){}}
\thicklines
\put(8,111){\makebox(25,10){\textbf{Energy}}}
\put(43,111){\makebox(42,10){\textbf{Fields}}}
\put(101,111){\makebox(52,10){\textbf{Effective Theory}}}
\put(5,110){\line(1,0){146}} {
\put(8,76){\makebox(25,30){$M_W$}}
\put(47,78){\framebox(34,25){
   $\ba W, Z, \gamma, G_a \\  \tau, \mu, e, \nu_i \\ t, b, c, s, d, u \ea $}}
\put(101,76){\makebox(52,30){Standard Model}}

\put(8,38){\makebox(25,20){$\lsim m_c$}}
\put(47,40){\framebox(34,16){
 $\ba  \gamma, G_a  \, ;\, \mu ,  e, \nu_i \\ s, d, u \ea $}}
\put(101,38){\makebox(52,20){$\cL_{\mathrm{QCD}}^{N_f=3}$,
             $\cL_{\mathrm{eff}}^{\Delta S=1,2}$}}

\put(8,0){\makebox(25,20){$M_K$}}
\put(47,2){\framebox(34,16){ 
 $\ba\gamma \; ;\; \mu , e, \nu_i  \\  \pi, K,\eta  \ea $}}
\put(101,0){\makebox(52,20){$\chi$PT}}
\linethickness{0.3mm}
\put(64,36){\vector(0,-1){15}}
\put(64,74){\vector(0,-1){15}}
\put(69,65){OPE}
\put(69,27){$N_C\to\infty $}}    
\end{picture}
}
\vskip -.5cm\mbox{}
\caption{Evolution from $M_W$ to the kaon mass scale.
  \label{fig:eff_th}}
\end{figure}
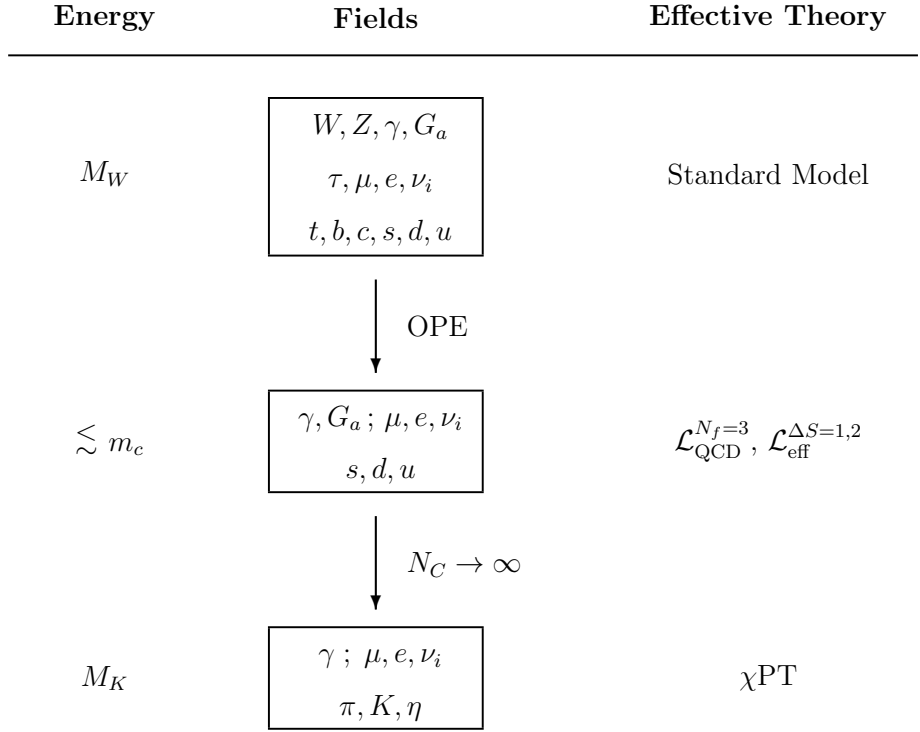 

In Figure~\ref{fig:eff_th}, we show schematically how the flavour-changing transitions are described at two different energy scales: at short distances one employs the effective $\Delta S = 1$ Lagrangian given by Eq.~(\ref{eq:shortLagr}), while at very low energies the $\chi$PT formalism introduced in Section~\ref{sec:EFTdesc} is more appropriate. The short-distance Lagrangian can only be used at scales where perturbation theory is well-defined, {\it i.e.}, $\mu_{\rm SD}\gtrsim 1\:$GeV. On the other hand, the chiral framework is valid in the non-perturbative regime, where all the fields of the heavy particles have been integrated out, but paying the price  of having a large number of unknown $\chi$PT couplings. These LECs must be determined either from data or using theoretical considerations. In the latter case, one needs to match both EFTs in a common region of validity. Unfortunately, performing consistently this non-perturbative matching is still very challenging~\cite{Gisbert:2017vvj,Bai:2015nea,Blum:2015ywa}.
 However, in the limit of a large number $N_C$ of QCD colours, the T-product of two colour-singlet quark currents factorizes and, since the quark currents have a well-known representation in terms of the Nambu-Goldstone bosons, one can make this matching at leading order in an expansion in powers of $1/N_C$. As a result, we obtain the electroweak chiral couplings ($g_8,\:g_{27},\:g_8\:g_{\rm ewk},\:g_8 N_i,\:g_{27} D_i,\:g_8 Z_i$) in terms of the strong and electromagnetic LECs of $\cO(p^n)$ with $n=2,4,6$ and $\cO(e^2 p^2)$, respectively.

\subsection[Weak couplings at $\cO(G_F p^2)$ and $\cO(e^2 G_8 p^0)$]{Weak couplings at $\boldsymbol{\cO(G_F p^2)}$ and $\boldsymbol{\cO(e^2 G_8 p^0)}$}\label{sec:LOgi}

At leading order in $1/N_C$, the chiral couplings of the nonleptonic electroweak Lagrangians of $\cO(G_F \, p^2)$ and $\cO(e^2 \, G_8 \, p^0)$, given by Eqs.~(\ref{eq:Lweak}) and (\ref{eq:Lelweak}), take the values \cite{Cirigliano:2003gt,Pallante:2001he}
\begin{align}
g_8^\infty\, &=\, -{2\over 5}\,C_1(\mu_{\rm SD})+{3\over 5}\,
C_2(\mu_{\rm SD})+C_4(\mu_{\rm SD})
- 16\, L_5 \, B(\mu_{\rm SD})\, C_6(\mu_{\rm SD})\, ,  \nl
g_{27}^\infty\, &=\, {3\over 5}\, \big[C_1(\mu_{\rm SD})+
C_2(\mu_{\rm SD})\big]\, , \label{eq:c2} \\
(e^2 g_8\, g_{\rm ewk})^\infty\, &= \, -3\, B(\mu_{\rm SD})\, 
C_8(\mu_{\rm SD}) 
- \frac{16}{3} \,  B(\mu_{\rm SD})\, C_6 (\mu_{\rm SD}) \, 
e^2 \, (K_9 - 2 K_{10})~, \nn 
\end{align}where
\begin{align} 
B(\mu_{\rm SD})\,\equiv\,
 \left[{M_K^2\over (m_s + m_d)(\mu_{\rm SD})\, F_\pi}\right]^2 
&\Biggl[ 1  -\fr{16 M_K^2}{F_\pi^2}\, (2 L_8-L_5)  
+ {8 M_\pi^2\over F_\pi^2}\,  L_5
\Biggr]~.
\label{eq:B0_comp}  
\end{align}
These large-$N_C$ expressions imply\footnote{The numerical inputs for $L_{5}$, $K_{9}$ and $K_{10}$ are presented below.}
\begin{align}
 g_8^\infty\:=\:&\bigg({{1.15^{\,+\,0.07}_{\,-\,0.12}}}_{\,\mu_{\rm SD}}\:\pm\:0.02_{\,L_{5,8}}\:\pm\:0.01_{\, m_s}\bigg)\nonumber\\[0.15cm]
&+\:\tau\:\bigg({0.78^{\,+\,0.09}_{\,-\,0.08}}_{\,\mu_{\rm SD}}\:\pm\:0.10_{\,L_{5,8}}\:\pm\:0.03_{\, m_s}\bigg)~,\label{eq:eq1}\\[0.25cm]
g_{27}^\infty\:=\:&0.46\:\pm\:0.01_{\,\mu_{\rm SD}}~,\label{eq:eq2}\\[0.25cm]
(g_8\:g_{\rm ewk})^\infty\:=\:&\bigg({{-1.57}^{\,+\, 1.00}_{\,-\,0.51}}_{\,\mu_{\rm SD}}\:\pm\:0.14_{\,L_{5,8}}\:\pm\:0.18_{\, K_i}\:\pm\:0.05_{\, m_s}\bigg)\nonumber\\[0.15cm]
&+\:\tau\:\bigg({{-20.4}^{\,+\, 1.6}_{\,-\,1.7}}_{\,\mu_{\rm SD}}\:\pm\:1.8_{\, L_{5,8}}\:\pm\:0.85_{\, K_i}\:\pm\:0.7_{\, m_s}\bigg)~,\label{eq:eq3}
\end{align}
where the first uncertainty has been estimated through the variation of the scale $\mu_{\rm SD}$ between 0.9 GeV and 1.2 GeV, while the second and third ones reflect the current errors on the strong LECs of $\cO(p^4)$ and the electromagnetic couplings of $\cO(e^2 p^2)$. The last error indicates the parametric uncertainty induced by the quark mass factor, which has been taken within the range $(m_s+m_d)(\mu_{\rm SD}=1\text{\:GeV})=131.8\pm 2.2\text{\:MeV}$.\footnote{Using as inputs the values of $\alpha_{s}(M_{Z})=0.11823 \pm 0.00081$, $m_{d} (N_{f}=3)=4.67 \pm 0.09 \, \mathrm{MeV}$ and $m_{s}(N_{f}=4)=93.44 \pm 0.68\, \mathrm{MeV}$ at $\mu_{\rm SD}= 2\, \mathrm{GeV}$, plus $\bar{m}_{Q}(\mu_{\rm SD}=\bar{m}_{Q})$ for the heavy quarks from \cite{Aoki:2019cca}, we use {\tt RunDec} \cite{Herren:2017osy} to decouple the fourth flavour ($m_{s}(N_{f}=3)=93.56 \pm 0.68\, \mathrm{MeV}$) and to obtain the quark masses at $1\,\mathrm{GeV}$, finding $m_{s}(\mu_{\rm SD}=1\, \mathrm{GeV})=125.6 \pm 0.9_{m_{s}} \pm 1.9_{\alpha_{s}} \, \mathrm{MeV}$ and $m_{d}(\mu_{\rm SD}=1\, \mathrm{GeV})=6.27 \pm 0.12_{m_{d}} \pm 0.09_{\alpha_{s}} \, \mathrm{MeV}$.}    Furthermore, we have computed the Wilson coefficients with two different definitions of $\gamma_5$ within dimensional regularisation, the Naive Dimensional Regularisation (NDR) and 't Hooft-Veltman (HV) \cite{tHooft:1972tcz} schemes, and have used an average of the two results. When computing physical amplitudes we have 
included a conservative error to account for this 
scheme dependence (see Appendix~\ref{app:results}).\footnote{With respect to Ref. \cite{Gisbert:2017vvj}, we have updated the values of the quark masses and the strong coupling, using inputs from Ref. \cite{Aoki:2019cca} and the recent ATLAS determination of the running top quark mass \cite{Aad:2019mkw}.} Notice that we take into account the full evolution from the electroweak scale to $\mu_{\rm SD}$, without any unnecessary expansion in powers of $1/N_C$; otherwise one would miss the large short-distance logarithms encoded in $C_i(\mu_{\rm SD})$ for $i\not= 6,8$. The large-$N_C$ approximation is only applied to the matching process between the short-distance and $\chi$PT descriptions.

The numerical results in Eqs.~(\ref{eq:eq1}) and (\ref{eq:eq2}) are quite far from their phenomenologically extracted values, including chiral loop corrections, $g_8\approx 3.6$ and $g_{27}\approx 0.29$ \cite{Cirigliano:2011ny}. This large deviation can be understood when one realizes how those operators that dominate the contributions to $g_{8}^\infty$ and $g_{27}^{\infty}$ have vanishing associated anomalous dimension in the large-$N_{C}$ limit. Relevant information on these anomalous dimensions that should be reflected in the hadronic matrix elements is then lost in this limit. This fact indicates the importance of $\cO(1/N_C)$ corrections in the CP-conserving amplitudes. 
Many efforts to estimate these contributions have been attempted in the past \cite{Bardeen:1986vz,Pich:1990mw,Donoghue:1993xd,Jamin:1994sv,Pich:1995qp,Antonelli:1995nv,Antonelli:1995gw,Bertolini:1997ir,Hambye:1998sma,Knecht:1998nn,Bijnens:1998ee,Donoghue:1999ku,Bijnens:2000im,Bertolini:2000dy,Narison:2000ys,Cirigliano:2001qw,Bijnens:2001ps,Knecht:2001bc,Cirigliano:2002jy,Hambye:2003cy,Buras:2014maa}, but a proper NLO matching in $1/N_C$ within a well-defined EFT framework is still lacking.
In Section~\ref{sec:fitChiPT}, we will perform a fit to $K\to\pi\pi$ data in order to obtain reliable predictions for the CP-conserving parts of $g_8$ and $g_{27}$.

Fortunately, this problem does not arise for the CP-odd contributions. The anomalous dimensions of the leading operators, $Q_{6}$ and $Q_{8}$, survive when $N_C\to\infty$, allowing us to keep track of all large logarithms.  Therefore, the $\chi$PT evaluation of both operators in the large-$N_{C}$ limit provides the correct dependence on the short-distance renormalization scale $\mu_{\rm SD}$,  given by $B(\mu_{\rm SD})\sim (1/(m_s + m_d)(\mu_{\rm SD}))^2\sim (\alpha_s(m_c) /\alpha_s(\mu_{\rm SD}))^{9/11} $, which exactly cancels the $\mu_{\rm SD}$ dependence of $C_{6,8}(\mu_{\rm SD})$ at large $N_C$. As a consequence, we have a much better control on the $\text{Im} A_I$ amplitudes, which makes their large-$N_C$ estimates 
more reliable than their CP-conserving counterparts. 

\subsection[Weak couplings at $\cO(G_F p^4)$ and $\cO(e^2 G_8 p^2)$]{Weak couplings at $\boldsymbol{\cO(G_F p^4)}$ and $\boldsymbol{\cO(e^2 G_8 p^2)}$}

At NLO, the large-$N_C$ matching fixes the couplings $G_8 N_i$, $G_{27} D_i$ and $G_8 Z_i$ of the non-leptonic weak and electroweak Lagrangians (\ref{eq:Lweak}) and (\ref{eq:Lelweak}). In this section, we compile the results obtained in Ref.~\cite{Cirigliano:2003gt}. Taking the definitions,
\begin{align}
\widetilde{C}_1 (\mu_{\rm SD}) &\equiv  
-\frac{2}{5} C_1 (\mu_{\rm SD}) + 
\frac{3}{5} C_2 (\mu_{\rm SD}) + C_4 (\mu_{\rm SD})\, ,\\[0.5cm]
\widetilde{C}_2 (\mu_{\rm SD}) &\equiv 
+\frac{3}{5} C_1 (\mu_{\rm SD}) 
-\frac{2}{5} C_2 (\mu_{\rm SD}) + C_3 (\mu_{\rm SD}) - C_5 (\mu_{\rm SD})\, ,
\end{align}
the non-vanishing LECs contributing to the $K\to\pi\pi$ amplitudes can be parametrized as follows:
\begin{eqnarray}
(g_{27}\, D_4)^\infty &=&  4\,  L_5 \, g_{27}^\infty\, ,  \label{eq:g27D4}
\end{eqnarray}
\begin{eqnarray}
(g_8 \, N_i)^\infty &=& n_i \, L_5 \, \widetilde{C}_1 (\mu_{\rm SD}) +\,\mathcal{X}_i\, B(\mu_{\rm SD}) \, C_6 (\mu_{\rm SD})   
\nonumber\\[0.3cm]
&=& n_i \, L_5\,\left(g_8^\infty \,+\,B(\mu_{\rm SD})\, C_6(\mu_{\rm SD})\,\left[16\,L_5\,+\,\frac{\mathcal{X}_i}{n_i\,L_5}\right] \right) ,
\label{eq:g8Ni}
\end{eqnarray}
with $n_i$ and $\mathcal{X}_i$ 
given
in Table~\ref{table:g8N_i} of Appendix~\ref{sec:largeNCparameters} as functions of the LECs of Eq. (\ref{eq:Lstrong}),
and
\begin{eqnarray}\label{eq:g8Zi-LargeNc}
(g_8 \, Z_i)^\infty &=& \mathcal{K}^{(1)}_i \, \widetilde{C}_1 (\mu_{\rm SD}) +\mathcal{K}^{(2)}_i \, \widetilde{C}_2 (\mu_{\rm SD}) + \,\mathcal{K}^{(3)}_i B(\mu_{\rm SD}) \, C_6 (\mu_{\rm SD})  
\\[0.3cm]
&+&\frac{1}{e^2}\,\Big\{\mathcal{K}^{(4)}_i C_7 (\mu_{\rm SD})+\mathcal{K}^{(5)}_i B(\mu_{\rm SD})\, C_8 (\mu_{\rm SD}) + \mathcal{K}^{(6)}_i C_9 (\mu_{\rm SD})+ \mathcal{K}^{(7)}_i C_{10} (\mu_{\rm SD})\Big\} ,\nonumber
\end{eqnarray}
where the constants $\mathcal{K}^{(k)}_i$ are given in Table~\ref{table:g8Z_i} of Appendix~\ref{sec:largeNCparameters}.

The dependence on the $\chi$PT renormalization scale $\nu_\chi$ is of $\cO(1/N_C)$ and, therefore, is absent from these large-$N_C$ expressions. To account for this systematic uncertainty, we will vary $\nu_\chi$ between 0.6 GeV and 1 GeV in the loop contributions and the resulting numerical fluctuations will be added as an additional error in the predicted amplitudes.

\subsection[Strong couplings of $\cO(p^4)$ and $\cO(p^6)$]{Strong couplings of $\boldsymbol{\cO(p^4)}$ and $\boldsymbol{\cO(p^6)}$}

The $K\to\pi\pi$ amplitudes have an explicit dependence on some LECs of the $\cO(p^4)$ strong Lagrangian, in the large-$N_C$ limit. We have already set $L_4^\infty=L_6^\infty=0$, which are rigorous QCD results at $N_C\to\infty$. The large-$N_C$ estimates based on resonance saturation are  known to provide an excellent description of the $L_i$ couplings at $\nu_\chi\sim M_\rho$ \cite{Pich:2002xy}. For the LECs that are relevant here, they give \cite{Ecker:1988te,Pich:2002xy}
\begin{equation}
 L_5^\infty = \frac{8}{3}\,  L_8^\infty = -4\, (2 L_8 - L_5)^\infty =
 \frac{F_\pi^2}{4 M_S^2}\approx 1.0\cdot 10^{-3}\, ,
 \label{eq:L_i_valuesnc}
\end{equation}
and
\begin{equation}\label{eq:L7_LargeNc}
L_7^\infty = -\frac{F_\pi^2}{48 M_{\eta_1}^2}\approx -0.27\cdot 10^{-3}\, ,
\end{equation}
with $F_\pi= 92.1$ MeV, $M_S\approx 1500$ MeV and $M_{\eta_1}= 804$ MeV \cite{Ecker:1988te}.
In Table~\ref{table:L578} we compare this numerical estimate with the LECs extracted from the most recent $\cO(p^4)$ and $\cO(p^6)$ $\chi$PT fits to kaon and pion data \cite{Bijnens:2014lea}, and with the values of $L_5^r(M_\rho)$ and $L_8^r(M_\rho)$ advocated in the current FLAG compilation of lattice results \cite{Aoki:2019cca}, which have been obtained by the HPQCD collaboration \cite{Dowdall:2013rya} analyzing the kaon and pion decay constants at different quark masses with $N_f=2+1+1$ dynamical flavours. 
All these determinations are in excellent agreement with the large-$N_C$ estimates.
Although much more precise, the  $\cO(p^6)$ $\chi$PT values of $L_5^r(M_\rho)$ and $L_8^r(M_\rho)$ are sensitive to assumptions concerning the $\cO(p^6)$ LECs. $L_7$ has not been yet extracted from lattice data but, fortunately, its $\chi$PT value remains very stable under different fit conditions. Note that $L_7$ does not depend on the $\chi$PT renormalization scale. In our numerical analysis, we will adopt the values:
\begin{equation}\label{eq:Li-inputs}
\begin{array}{ccc}
L_5^r(M_\rho) = (1.20\pm 0.10)\cdot 10^{-3}\, ,    
&\qquad & 
L_8^r(M_\rho) = (0.53\pm 0.11)\cdot 10^{-3}\, ,
\\
(2\: L_8^r - L_5^r)(M_\rho) = (-0.15\pm 0.20)\cdot 10^{-3}\, ,
&\qquad &
L_7 = (-0.32\pm 0.10)\cdot 10^{-3}\, .
\end{array}
\end{equation}

The chosen ranges for the nearly uncorrelated (in the different fits) LECs $L_5$ and $2L_{8}-L_{5}$ result from averaging the central lattice and $\cO(p^4)$ $\chi$PT values, rounding-up the uncertainties so that they are not smaller than the most precise value. $L_{8}$ is obtained from the previous two values, neglecting their small correlation. For $L_7$ we have applied  the same prescription to the $\cO(p^4)$ and $\cO(p^6)$ chiral results, but slightly rounding-up the $\cO(p^6)$ uncertainty.

\begin{table}[tb]
\begin{center}
\begin{tabular}{|c|c|c|c|c|}
\hline
& $L_5^r(M_\rho)$ & $L_8^r(M_\rho)$ & $(2 L_8^r - L_5^r)(M_\rho)$ & $L_7$
\\ \hline
Large-$N_C$ estimate & $1.0$ & $0.4$ & $-0.2$ & $-0.27$
\\
$\cO(p^4)$ $\chi$PT fit & $1.2\pm 0.1$ & $0.5\pm 0.2$ & $-0.2\pm 0.4$ & $-0.3\pm 0.2$
\\
$\cO(p^6)$ $\chi$PT fit & $1.01\pm 0.06$ & $0.47\pm 0.10$ & $-0.07\pm 0.18$ & $-0.34\pm 0.09$
\\
Lattice & $1.19\pm 0.25$ & $0.55\pm 0.15$ & $-0.10\pm 0.20$ & ---
\\ \hline
\end{tabular}
\end{center}
\caption{Comparison of the large-$N_C$ estimates for the relevant strong LECs of $\cO(p^4)$  \cite{Pich:2002xy} with the values extracted from $\cO(p^4)$ and $\cO(p^6)$ $\chi$PT fits \cite{Bijnens:2014lea} and the lattice results \cite{Aoki:2019cca,Dowdall:2013rya}. All numbers are given in units of $10^{-3}$.}
\label{table:L578}
\end{table}

The strong LECs of the $\cO(p^6)$ Lagrangian enter into the amplitudes through the coefficients $\mathcal{X}_i$ of Eq.~\eqref{eq:g8Ni}, which only depend on  $X_{12}$, $X_{14-20}$, $X_{31}$, $X_{33}$, $X_{34}$, $X_{37}$, $X_{38}$, $X_{91}$ and $X_{94}$. The dependence on $X_{37}$ and $X_{94}$ exactly cancels, however, in all $\Delta_C\cA^{(X)}_n$ amplitudes; thus these couplings are not needed. Using Resonance Chiral Theory (R$\chi$T) \cite{Ecker:1988te,Ecker:1989yg}, these LECs can be estimated in terms of meson resonance parameters, through the tree-level exchange of the lightest resonance states. This amounts to perform the matching between the $\chi$PT and R$\chi$T Lagrangians at leading order in $1/N_C$, in the single-resonance approximation. An analysis of all resonance contributions to the $X_i$ couplings can be found in Ref.~\cite{Cirigliano:2006hb}. Furthermore, a complete analysis of the $\eta_1$ contributions to the chiral low-energy constants of $\cO(p^6)$ was presented in Ref.~\cite{Kaiser:2007zz}. Combining both results, we obtain the values given in Table~\ref{table:X_i}. 

As expected for the $K\to\pi\pi$ amplitudes, the relevant couplings do not receive contributions from vector and axial-vector exchanges. Moreover, all $\eta_1$ contributions coming from the $\widetilde{X}^{\eta_1}_{i}$ factors in Table~\ref{table:X_i} cancel also in the combinations $\mathcal{X}_i$ that govern the $(g_8 N_i)^\infty$ LECs (see Appendix~\ref{sec:largeNCparameters}), as it should. 
The exchange of $\eta_1$ mesons can only contribute indirectly to $K\to\pi\pi$, through the dependence on $L_7$ of the $\pi^0-\eta$ mixing correction $\varepsilon^{(4)}_S$ in Eq.~\eqn{eq:epsilon4S}, which gives rise to the  term proportional to $L_7 L_8$ in $\mathcal{X}_{13}$. This unique $\eta_1$ contribution appears in the NLO local corrections $\Delta_C\mathcal{A}_{1/2,3/2}^{(\varepsilon)}$ and represents one of the largest sources of uncertainty in our numerical results.

\begin{table}[tb]
\begin{center}
\begin{tabular}{|c|c|}
\hline
$X_i/F^2$ & Large-$N_C$ prediction   \\ \hline
12 &  $-\:\frac{c_d\:c_m}{2\:M_S^4}$  \\ \hline
14 &  $-\:\frac{d_m^2}{4\:M_P^4}\:+\:(\bar{\lambda}^{SS}_1)'\:+\:2\:\frac{c_d}{c_m}\:(\bar{\lambda}^{SS}_3)'$  \\ \hline
15 &  $0$  \\ \hline
16 &  $0$  \\ \hline
17 &  $-\:\frac{d_m^2}{4\:M_P^4}\:+\:\bar{\lambda}^{SS}_2$  \\ \hline
18 &  $\widetilde{X}_{18}^{\eta_1}$  \\ \hline
19 &  $\frac{c_d\:c_m}{27\:M_S^4}\:+\:
\frac{\bar{\lambda}_4^S}{9}\:+
(\bar{\lambda}_3^{SS})'
\:+\: \widetilde{X}_{19}^{\eta_1}$  \\ \hline
20 & $-\:\frac{c_d\:c_m}{18\:M_S^4}
\:-\:\frac{\bar{\lambda}_4^S}{6}
\:+\:\widetilde{X}_{20}^{\eta_1}$   \\ \hline
31 &  $-\:\frac{d_m^2}{2\:M_P^4}\:-\:\frac{7}{18}\:\frac{c_d\:c_m}{M_S^4}
\:+\:\frac{\bar{\lambda}_4^S}{3}
\:-\:2\:(\bar{\lambda}^{SP}_2)'\:+\: \widetilde{X}_{31}^{\eta_1}$  \\ \hline
33 &  $\frac{d_m^2}{6\:M_P^4}\:+\:\frac{2}{9}\:\frac{c_d\:c_m}{M_S^4}
\:+\:\frac{\bar{\lambda}_4^S}{6}\:+\:\bar{\lambda}_5^S\:-\:\bar{\lambda}^{P}_3 
\:+\:\widetilde{X}_{33}^{\eta_1}$  \\ \hline
34 &  $\frac{d_m^2}{2\:M_P^4}\:+\:\frac{c_d\:c_m}{2\:M_S^4}\:+\:\frac{c_m^2}{2\:M_S^4}\:-\:\frac{d_m^2}{M_P^2\:M_S^2}$  \\ \hline
38 & $-\:\frac{d_m^2}{2\:M_P^4}\:+\:\frac{c_m^2}{2\:M_S^4}$   \\ \hline
91 &  $2\:\frac{d_m^2}{M_P^4}$  \\ \hline
\end{tabular}
\end{center}
\caption{Large-$N_C$ predictions for the relevant strong LECs of $\cO(p^6)$, in $F^2$ units  \cite{Cirigliano:2006hb}.}
\label{table:X_i}
\end{table}

Thus, only contributions from scalar and pseudoscalar resonance-exchange enter into the relevant $X_i$ LECs in Table~\ref{table:X_i}. 
The LO R$\chi$T couplings have been determined within the single-resonance approximation, which gives the relations \cite{Pich:2002xy}:
\begin{align}
c_m=c_d=\sqrt{2}\: d_m\:=\:F_\pi/2\, ,
\qquad\qquad 
M_P=\sqrt{2}\:M_S\, .
\label{eq:SRA}
\end{align}
These couplings correspond to $\cO(p^2)$ chiral structures with Goldstone fields coupled to a single resonance multiplet, either scalar ($c_{d,m}$) or pseudoscalar ($d_m$). The table contains, in addition, contributions from $\cO(p^4)$ chiral structures with one resonance ($\bar\lambda_i^R$) and $\cO(p^2)$ terms with two resonances ($\bar\lambda_i^{RR'}$) that are currently unknown. We are only aware of one estimate of $\lambda_3^{SS} \equiv \bar\lambda_3^{SS} M_S^4/c_m^2$, determined from the scalar resonance spectrum~\cite{Cirigliano:2003yq}, which we update in Appendix~\ref{app:lambda3}. We obtain:
\begin{equation}
M_S\, =\, 1478\:\text{MeV}\, ,
\qquad\qquad
\lambda_3^{SS}\,=\, 0.1548\, .
\label{eq:massscalar}
\end{equation}

In the absence of better information, we will take null values for the unknown $\bar\lambda_i^R$ and $\bar\lambda_i^{RR'}$ couplings.
In order to estimate the size of uncertainties in any observable $F$ associated to the LECs $X_{i}$, we will take:
\begin{align}\label{eq:XiError}
\text{error of }F\; =\;\frac{|F(X_i)-F(0)|}{N_C}\, .    
\end{align}

\subsection[Electromagnetic couplings of $\cO(e^2 p^2)$]{Electromagnetic couplings of $\boldsymbol{\cO(e^2 p^2)}$}

The electromagnetic LECs $K_i$ can be expressed as convolutions of QCD correlators with a photon propagator \cite{Moussallam:1997xx}, and their evaluation involves an integration over the virtual photon momenta. Therefore, they have an explicit dependence on the $\chi$PT renormalization scale $\nu_\chi$, already at leading order in $1/N_C$. In Ref. \cite{Ananthanarayan:2004qk}, the couplings $K_{1-6}^r$ have been estimated by computing 4-point Green functions (two currents and two electromagnetic spurion fields) in $\chi$PT  and matching them with their R$\chi$T estimates (neglecting pseudoscalar contributions). The R$\chi$T  couplings are obtained by imposing short-distance constraints. They find 
\begin{align}
 K_{1}^{r}(M_\rho)\, &=\, -K_3^r(M_\rho)\, =\, -2.71\,\cdot 10^{-3}\, ,
 \qquad \qquad
 K_{5}^{r}(M_\rho)\, =\, 11.59\,\cdot 10^{-3}\, ,
 \nonumber\\
 K_{2}^{r}(M_\rho)\, &=\, \frac{1}{2}\, K_4^r(M_\rho)\, =\,0.69\,\cdot 10^{-3}\, ,
 \qquad\qquad\;\;\, 
 K_{6}^{r}(M_\rho)\, =\, 2.77\,\cdot 10^{-3}\, .
\end{align}

The remaining couplings can be accessed through the study of two- and three-point functions. $K_{7,8}^r$
turn out to be $1/N_C$ suppressed, {\it i.e.}, $K_7^r(M_\rho)\approx K_7^r(M_\rho)\approx 0$ \cite{Moussallam:1997xx}. $K^r_{9-13}$ are gauge dependent, while $K^r_{9-12}$ depend also on the short-distance renormalization scale $\mu_{\rm SD}$. Those dependences cancel with the photon loop contributions in the physical decay amplitudes. The explicit values we quote below refer to the Feynman gauge ($\xi=1$) and $\mu_{\rm SD}=1$~GeV \cite{Moussallam:1997xx,Ananthanarayan:2004qk,Cirigliano:2003gt,Bijnens:1996kk,Albaladejo:2017hhj}:
\begin{equation}
K^{r}_9(M_\rho)\, =\, 2.2 \cdot 10^{-3}\, ,
\qquad\qquad
K^r_{10}(M_\rho)\, =\, 6.5 \cdot 10^{-3}\, ,
\nonumber\end{equation}\begin{equation}
K^r_{11}(M_\rho)\, =\, 1.26 \cdot 10^{-3}\, ,
\qquad
K^r_{12}(M_\rho)\, =\, -4.2 \cdot 10^{-3}\, ,
\qquad
K^r_{13}(M_\rho)\, =\, 4.7 \cdot 10^{-3}\, .
\end{equation}

The uncertainties associated with these LECs will be also estimated following the method indicated in Eq.~\eqn{eq:XiError}.

\section{Anatomy of isospin-breaking parameters in $\boldsymbol{\epsilon^\prime}$  }\label{sec:Decomp}

At first order in isospin corrections, Eq.~(\ref{eq:eps1}) can be written as~\cite{Cirigliano:2003gt,Cirigliano:2003nn}
\begin{equation} 
\epsilon' = - \frac{i}{\sqrt{2}} \, e^{i ( \chi_2 - \chi_0 )} \, 
\omega_+  \,   \left[ 
\frac{\imag A_{0}^{(0)}}{ \real A_{0}^{(0)}} \, 
(1 + \Delta_0 + f_{5/2}) - \frac{\imag A_{2}}{ \real
  A_{2}^{(0)}} \right] , 
\label{eq:cpiso}
\end{equation}
where the superscript $(0)$ denotes the isospin limit, and the different sources of isospin-breaking effects are made explicit.
From the measured $K^+\to\pi^+\pi^0$ and $K^0\to\pi\pi$ rates, one actually determines the ratio
\begin{equation}\label{eq:omegaplus}
\omega_+ \, =\, \frac{\real A_{2}^{+}}{\real A_{0} }
\, =\, \omega \left\{ 1 + f_{5/2}\right\} ,
\end{equation}
which differs from $\omega = {\real A_{2}}/{\real A_{0}}$ by the small electromagnetic correction $f_{5/2}$. The breaking of isospin in the leading $I=0$ amplitude is parametrized through
\begin{equation}
\Delta_0 \, = \, \frac{\imag A_0}{\imag A_0^{(0)}} \,
\frac{\real A_0^{(0)}}{\real A_0}  - 1  \, ,
\end{equation}
while we can approximate ${ \real A_{2}}\approx { \real A_{2}^{(0)}}$ because ${\imag A_{2}}$ is already an isospin-breaking correction.

In order to determine these corrections, it is useful to write the CP-violating amplitudes as
\begin{eqnarray}  
A_0 \,e^{i \chi_0 } &=&   \cA_{1/2}^{(0)} + \delta \cA_{1/2} ,
\nn \\
 A_2 \,e^{i \chi_2 }&=& \cA_{3/2}^{(0)} + \delta \cA_{3/2}+ 
\cA_{5/2} ~,
\label{eq:A02}   
\end{eqnarray}
where $\delta \cA_{1/2,3/2}$ and $\cA_{5/2}$ are first order in isospin
violation. 
The amplitudes $\cA_{\Delta I}$ have both
absorptive ($\text{Abs} ~\cA_{\Delta I}$) and dispersive ($\text{Disp} ~\cA_{\Delta I}$) parts. Therefore, the loop-induced phases $\chi_{I}$ have to be carefully separated from the CP-violating ones. Expanding to first order in CP and isospin violation, one finds \cite{Cirigliano:2003gt}:
\begin{eqnarray}  
\imag A_0^{(0)} &=&  \left| \cA_{1/2}^{(0)} \right|^{-1} \, 
\left\{\imag [ \text{Disp} \,\cA_{1/2}^{(0)} ] \,
\real [ \text{Disp} \, \cA_{1/2}^{(0)} ] + 
\imag [ \text{Abs} \,\cA_{1/2}^{(0)}]  \,
\real[ \text{Abs} \,  \cA_{1/2}^{(0)} ] \right\}  
\label{eq:ima00} \, ,\\
\imag A_2 \ \,  &=&  \left| \cA_{3/2}^{(0)} \right|^{-1} \, 
\left\{\imag [ \text{Disp} \, 
\left( \delta \cA_{3/2} + \cA_{5/2} \right) ]  \,
\real [ \text{Disp} \, \cA_{3/2}^{(0)} ] \right.  \nn \\
&& \left. \qquad \qquad + \   
\imag [ \text{Abs} \left( \delta  \cA_{3/2} + \cA_{5/2} \right) ]  \,
\real[ \text{Abs} \,  \cA_{3/2}^{(0)} ] \right\}  \, , \\
\Delta_0 &=& -2 \left| \cA_{1/2}^{(0)} \right|^{-2} \, 
\left( \real  [ \text{Disp} \, \cA_{1/2}^{(0)} ]~ 
\real [\text{Disp} \, \delta \cA_{1/2}] + 
\real [ \text{Abs} \,\cA_{1/2}^{(0)}]~ 
\real [\text{Abs} \, \delta \cA_{1/2}]
\right) \quad\nn  \\
&+& \left[ \imag [ \text{Disp} \,\cA_{1/2}^{(0)} ] \,
\real [ \text{Disp} \, \cA_{1/2}^{(0)} ] + 
\imag [ \text{Abs} \,\cA_{1/2}^{(0)}]  \,
\real[ \text{Abs} \,  \cA_{1/2}^{(0)} ] \right]^{-1}  
\nn  \\ 
&& \times\,\left\{\imag [ \text{Disp} \,\delta\cA_{1/2} ] \,
\real [ \text{Disp} \,\cA_{1/2}^{(0)} ] + 
\imag [ \text{Disp} \,\cA_{1/2}^{(0)} ] \,
\real[ \text{Disp} \,  \delta\cA_{1/2} ] \right. \nn \\
&&  \hskip .4cm\mbox{} +  \,  \left. \imag [  \text{Abs} \,\delta\cA_{1/2} ] \,
\real[ \text{Abs} \,  \cA_{1/2}^{(0)} ] + 
\imag [ \text{Abs} \,\cA_{1/2}^{(0)} ] \,
\real [ \text{Abs} \, \delta\cA_{1/2} ] \right\}\, , \\ 
f_{5/2} &=& \frac{5}{3} \left| \cA_{3/2}^{(0)} \right|^{-2} \, 
\left\{\real [ \text{Disp} \,\cA_{3/2}^{(0)} ] \,
\real [ \text{Disp} \, \cA_{5/2} ] + 
\real [ \text{Abs} \,\cA_{3/2}^{(0)}]  \,
\real[ \text{Abs} \,  \cA_{5/2} ] \right\}
\label{eq:f52}.   
\end{eqnarray} 

It is convenient to separate the electroweak penguin contribution to $\imag A_2$ from the
isospin-breaking effects generated by other four-quark operators: 
\begin{equation} 
\imag A_2 = \imag A_2^{\rm emp} \ + \ \imag  A_2^{\rm non-emp} \ .  
\end{equation}  
This separation depends on the renormalization scheme,\footnote{Only the electromagnetic contribution is scheme dependent. We use the $\overline{\mathrm{MS}}$ scheme with both NDR and HV prescriptions, assigning an extra uncertainty due to the very small resulting differences.} but allows one to identify the terms that are enhanced by the ratio $1/\omega$ and write them explicitly as corrections to the $I=0$ side through the parameter
\begin{equation}
\Omega_{\rm IB} \,  =\, \displaystyle\frac{\real A_0^{(0)} }
{ \real A_2^{(0)} } \cdot \displaystyle\frac{\imag A_2^{\rm non-emp} }
{ \imag A_0^{(0)} } \, .
\end{equation}
The splitting is easily performed at leading order in $1/N_C$ through the matching procedure between the short-distance and $\chi$PT descriptions. The electroweak LECs in $\imag A_2^{\rm non-emp}$ are calculated by setting to zero the Wilson coefficients $C_{7-10}$ of the electroweak penguin operators. We can then write $\epsp$ as
\begin{equation}  
\epsp = - \displaystyle\frac{i}{\sqrt{2}} \, e^{i ( \chi_2 - \chi_0 )} \, 
\omega_+  \,   \left[ 
\displaystyle\frac{\imag A_{0}^{(0)} }{ \real A_{0}^{(0)} } \, 
(1 - \Omega_{\rm eff}) - \displaystyle\frac{\imag A_{2}^{\rm emp}}{ \real
  A_{2}^{(0)} } \right]  ,
\label{eq:cpeff}
\end{equation} 
with
\begin{equation}  
\Omega_{\rm eff}\, =\, \Omega_{\rm IB} - \Delta_0 - f_{5/2} \, .  
\label{eq:omegaeff}
\end{equation} 

\section{Numerical results}
\label{sec:numbers}

At this point, we have all the theoretical ingredients to provide a numerical prediction for the isospin-breaking effects in $K\to \pi\pi$. In the following subsections, we present each of the numerical results that enter in the estimation of these corrections. 

\subsection{Amplitudes at NLO}\label{sec:ampnlonum}
In this subsection, we present the numerical results of the different isospin amplitudes, $\mathcal{A}_n$ with $n=1/2,\:3/2$ and $5/2$. Tables~\ref{tab:amp12}, \ref{tab:amp32} and \ref{tab:amp52}, which supersede Tables 1, 2 and 3 of Ref.~\cite{Cirigliano:2003gt}, display the following information:

\begin{table}[p]
\begin{center}
$
\begin{array}{|c|c|c|c|c|}\hline
(X) & a_{1/2}^{(X)} &\Delta_{L} \cA_{1/2}^{(X)}  &
[\Delta_{C} \cA_{1/2}^{(X)}]^+   & 
[\Delta_{C} \cA_{1/2}^{(X)}]^-  \\ \hline
 27 & \frac{\sqrt{2}}{9} & 1.03+0.47\: i & 0.01\,{}^{+0.00}_{-0.00}\, {}^{+0.65}_{-0.62} & 0.01\, {}^{+0.00}_{-0.00}\, {}^{+0.65}_{-0.62} \\ \hline
 8 & \sqrt{2} & 0.27+0.47 \:i & 0.02\,{}^{+0.00}_{-0.00}\, {}^{+0.05}_{-0.05} & 0.10\,{}^{+0.00}_{-0.00}\, {}^{+0.05}_{-0.05} \\ \hline
 \varepsilon & -\frac{2\sqrt{2}}{3\sqrt{3}} & 0.26+0.47\:   i & -0.37\,{}^{+0.04}_{-0.10}\,{}^{+0.05}_{-0.06} & 1.39\, {}^{+0.02}_{-0.02} \,{}^{+0.05}_{-0.06} \\ \hline
 \gamma & \text{-} & -1.39 &
   -0.47\,{}^{+0.18}_{-0.08}\,{}^{+0.26}_{-0.27}& -10.67\, {}^{+0.88}_{-0.81}\, {}^{+0.26}_{-0.27} \\ \hline
 \text{Z} & \frac{4\sqrt{2}}{3} & -1.07+0.80\: i & -0.11\,{}^{+0.00}_{-0.01}
 \,{}^{+0.17}_{-0.18} & 0.13\, {}^{+0.00}_{-0.00}
 \,{}^{+0.17}_{-0.18} \\ \hline
 \text{g} & \frac{2\sqrt{2}}{3} & 0.28+0.47\: i & -0.19\, {}^{+0.00}_{-0.00} \,{}^{+0.01}_{-0.01}  & -0.19\, {}^{+0.00}_{-0.00}\, {}^{+0.01}_{-0.01} \\ \hline
\end{array}
$
\end{center}
\caption{NLO  loop and local counterterm  amplitudes $\mathcal{A}_{1/2}$.
The two uncertainties in the local amplitudes are associated with the 
variations of the short-distance scale $\mu_{\rm SD}$ and the chiral scale $\nu_\chi$, respectively.}
\label{tab:amp12}

\begin{center}
$
\begin{array}{|c|c|c|c|c|}\hline
(X) & a_{3/2}^{(X)} &\Delta_{L} \cA_{3/2}^{(X)}  &
[\Delta_{C} \cA_{3/2}^{(X)}]^+   & 
[\Delta_{C} \cA_{3/2}^{(X)}]^-  \\ \hline
 27 & \frac{10}{9} & -0.04-0.21\: i & 0.01\,{}^{+0.00}_{-0.00}\, {}^{+0.05}_{-0.05} & 0.01\,{}^{+0.00}_{-0.00}\, {}^{+0.05}_{-0.05}  \\ \hline
 \varepsilon & \frac{4}{3\sqrt{3}} & -0.70-0.21\:
   i & -0.35\, {}^{+0.04}_{-0.11}\, {}^{+0.48}_{-0.50} & 1.50\, {}^{+0.02}_{-0.02} \,{}^{+0.48}_{-0.50}  \\ \hline
 \gamma & \text{-} & -0.47
   & 0.40\,{}^{+0.09}_{-0.04}\, {}^{+0.08}_{-0.09} & -0.09\,{}^{+0.14}_{-0.10} \,{}^{+0.08}_{-0.09}  \\ \hline
 \text{Z} & \frac{4}{3} & -0.87-0.79\: i & 0.01\,{}^{+0.00}_{-0.00}\,{}^{+0.32}_{-0.33} & 0.07 \,{}^{+0.00}_{-0.00}\,{}^{+0.32}_{-0.33}  \\ \hline
 \text{g} & \frac{2}{3} & -0.50-0.21\: i & -0.19\,{}^{+0.00}_{-0.00}\,{}^{+0.19}_{-0.20}  & -0.19\,{}^{+0.00}_{-0.00}\,{}^{+0.19}_{-0.20}  \\ \hline
\end{array}$
\end{center}
\caption{NLO loop and local counterterm amplitudes $\mathcal{A}_{3/2}$.
The two uncertainties in the local amplitudes are associated with the 
variations of the short-distance scale $\mu_{\rm SD}$ and the chiral scale $\nu_\chi$, respectively.}
\label{tab:amp32}

\begin{center}
$
\begin{array}{|c|c|c|c|c|} \hline
(X) & a_{5/2}^{(X)} &\Delta_{L} \cA_{5/2}^{(X)}  &
[\Delta_{C} \cA_{5/2}^{(X)}]^+   & 
[\Delta_{C} \cA_{5/2}^{(X)}]^-   \\ \hline
 \gamma & \text{-} & -0.51
   & -0.15\,{}^{+0.02}_{-0.01}\,{}^{+0.10}_{-0.11} & -0.54\,{}^{+0.00}_{-0.00} \,{}^{+0.10}_{-0.11} \\ \hline
 \text{Z} & \text{-} & -0.93-1.16\: i & -0.17\,{}^{+0.01}_{-0.01}\,{}^{+0.41}_{-0.43} &
   0.09\,{}^{+0.00}_{-0.00}\,{}^{+0.41}_{-0.43}  \\ \hline
\end{array}$
\end{center}
\caption{NLO loop and local counterterm amplitudes $\mathcal{A}_{5/2}$.
The two uncertainties in the local amplitudes are associated with the 
variations of the short-distance scale $\mu_{\rm SD}$ and the chiral scale $\nu_\chi$, respectively.}
\label{tab:amp52}
\end{table}
\begin{itemize}
    \item The type of contribution $(X)$ in the first column.
    \item The LO contributions $a_n^{(X)}$ in the second column.
    \item The NLO loop contributions $\Delta_{L} \cA_{n}^{(X)}$,  with the absorptive and dispersive components, in the third column. Absorptive contributions are independent on the chiral renormalization scale $\nu_\chi$. For the dispersive contributions, $\nu_\chi$ is fixed to 0.77 GeV.
    \item The NLO local corrections to the CP-even and CP-odd amplitudes, $[\Delta_{C} \cA_{n}^{(X)}]^+$ and $[\Delta_{C} \cA_{n}^{(X)}]^-$ respectively in the last columns, where
\begin{equation}
[\Delta_{C} \cA_{n}^{(X)}]^\pm \ = \ \left\{ 
\begin{array}{ccc} 
\displaystyle\frac{ {\real \atop \imag} \left( G_{27} \ \Delta_{C} \cA_{n}^{(27)} 
\right)}{  {\real \atop \imag}( G_{27} )}  &   &  X = 27,    \\[0.6cm] 
\displaystyle\frac{ {\real \atop \imag} \left( G_8 g_{\rm ewk} \ \Delta_{C} \cA_{n}^{(g)}
 \right)}{ {\real \atop \imag} ( G_8 g_{\rm ewk} )}  &   &  X = g,   \\[0.6cm] 
\displaystyle\frac{ {\real \atop \imag}  \left( G_8 \ \Delta_{C} \cA_{n}^{(X)} \right)}{ 
{\real \atop \imag} ( G_8 )}  &   &  X = 8, Z, \varepsilon, \gamma~.    
\end{array}
\right. 
\end{equation}
The estimation of NLO local contributions represents the main uncertainty in our results. In Tables~\ref{tab:amp12}, \ref{tab:amp32} and \ref{tab:amp52}, we quote two different sources of uncertainties. The first error is related with the lack of cancellation of the short-distance scale $\mu_{\rm SD}$. We estimate it by varying this scale from 0.9 GeV to 1.2 GeV. The second error is associated to the missed logarithmic corrections due to applying the large-$N_C$ limit. In order to estimate them, we vary the chiral renormalization scale between 0.6 and 1 GeV. In most of the cases, this non-perturbative error dominates over the first one. The various LECs have been set to their central values.
\end{itemize}
The numerical results displayed in the tables are in good agreement with the findings of Ref.~\cite{Cirigliano:2003gt}. While the underlying physics behind the large values of  $\Delta_{L} \cA_{1/2,3/2}^{(Z)}$ and $[\Delta_{C} \cA_{1/2}^{(\gamma)}]^-$ is well understood  (related to the absorptive cut in the amplitudes), the larger than expected $[\Delta_{C} \cA_{1/2,3/2}^{(\varepsilon)}]^-$ values, very sensitive to the $L_{7}$ input, are not. It might be consequence of a numerical accident. While the size of the couplings $g_8 N^r_{i}$ is not larger than expected, their role appears enhanced in the amplitudes with large numerical prefactors.

\subsection[$\chi$PT fit to $K\to\pi\pi$ data]{$\boldsymbol{\chi}$PT fit to $\boldsymbol{K\to\pi\pi}$ data}\label{sec:fitChiPT}

In subsection~\ref{sec:LOgi}, we have seen the price of taking the large-$N_ C$ limit in the CP-even sector, reflected in an unphysical short-distance scale dependence for the observables. The large-$N_C$ estimate is unable to correctly predict the CP-conserving parts of $g_8$ and $g_{27}$. However, one can fit them to data. 
Since we include electromagnetic effects to first order in $\alpha$, we must consider the inclusive sum of the $K\to\pi\pi$ and $K\to\pi\pi\gamma$ decay rates. We denote by $\Gamma_n$
with $n=+-,00,+0$ the corresponding observable widths into the different $\pi\pi$ final states
and define the ratios \cite{Cirigliano:2003gt}
\begin{align}
C_n\:=\:\left(\frac{2\:\sqrt{s_n}\:\Gamma_n}{\widetilde{G}_n\:\Phi_n}\right)^{1/2},
\end{align}
where $\sqrt{s_n}$ is the center-of-mass energy (the physical kaon mass) and $\Phi_n$ the appropriate two-body phase space. The infrared-finite factors $\widetilde{G}_n = 1 + \cO(\alpha)$, which take into account the inclusive sum of virtual and real photons, are given in Ref.~\cite{Cirigliano:2003gt}. The quantities $C_n$ are directly related to the isospin amplitudes defined in Eq.~\eqref{eq:isodecomp}:
\begin{align}
&A_2^+\:=\:\frac{2}{3}\:C_{+0}\, ,\qquad\qquad (A_0)^2\:+\:(A_2)^2\:=\:\frac{2}{3}\:C_{+-}^2\:+\:\frac{1}{3}\:C_{00}^2\, ,
\nonumber\\[0.01cm]
&\frac{A_2}{A_0}\:\cos(\chi_0-\chi_2)\:=\:\frac{r-1+(\frac{A_2}{A_0})^2(2\: r-\frac{1}{2})}{\sqrt{2}\, (1+2\: r)}\, ,\label{eq:fitequations}
\end{align}
where  $r\equiv(C_{+-}/C_{00})^2$.

Extracting the $C_n$ factors from the measured partial widths $\Gamma_{+-,00,+0}$ \cite{PhysRevD.98.030001} and using the $\chi$PT representation of the $A_I$ amplitudes,
we can perform a fit to $g_8$, $g_{27}$ and the phase difference $\chi_0-\chi_2$.
We leave $\chi_0-\chi_2$ as an additional free parameter to be determined by the fit because an accurate $\chi \mathrm{PT}$ prediction of the phase-shift difference would require the inclusion of higher-loop corrections \cite{Colangelo:2001df,Cirigliano:2009rr}.

Assuming isospin conservation, we obtain the results shown in Table~\ref{tab:ic-lo-nlo}, from LO and NLO fits. The values of $\text{Re}\,A_0$, $\text{Re}\,A_2$ and $\chi_0-\chi_2$ are directly determined from the $C_n$ ratios and, therefore, are the same in both fits. The first errors originate in the experimental inputs, while the second ones in $g_8$ and $g_{27}$ reflect the sensitivity to the $\chi$PT scale $\nu_\chi$.  The octet coupling is also sensitive to the short-distance renormalization scale $\mu_{\rm SD}$ (third error). One observes a sizeable difference between the LO and NLO fitted values of $g_8$, while $g_{27}$ remains stable. This just illustrates the much larger size of the chiral loop corrections to the octet amplitude. Since the $\cO(p^4)$ corrections are positive (negative) in the octet (27) amplitude, the extracted value of $g_8$ ($g_{27}$) decreases (slightly increases) at NLO.

\begin{table}[th]
\begin{center}
\begin{tabular}{|c|c|c|}\hline
     & LO fit & NLO fit\\ \hline
$\real \:g_8$ & $4.985 \pm  0.002_{\rm exp}$ 
& $3.601 \pm 0.001_{\rm exp}\, {^{\,+\,0.139}_{\,-\,0.135}}_{\,\nu_\chi} \, {^{\,+\,0.010}_{\,-\,0.004}}_{\,\mu_{\rm SD}}$
\\\hline
$\real \:g_{27}$ & $0.286 \pm  0.001_{\rm exp}$ 
& $0.288 \pm 0.001_{\rm exp} \pm 0.014_{\,\nu_\chi} $
\\\hline
 $\chi_0-\chi_2$ & \multicolumn{2}{|c|}{$(44.78 \pm  0.98_{\rm exp})^\circ$}
 \\\hline
 \hline
$\real \:A_0 $
& \multicolumn{2}{|c|}{$(2.711 \pm  0.001_{\rm exp}) \cdot 10^{-7}\:\text{GeV}$}
\\\hline
$\real \:A_2 $ & \multicolumn{2}{|c|}{
$(1.212 \pm  0.003_{\rm exp}) \cdot 10^{-8}\:\text{GeV}$
}
\\\hline
$\real \:A_0/\real \:A_2$  &   
\multicolumn{2}{|c|}{
$22.36 \pm  0.05_{\rm exp}$
}
\\\hline
\end{tabular}
\caption{\label{tab:ic-lo-nlo} LO and NLO fits to the $K\to\pi\pi$ amplitudes in the limit of isospin conservation.}
\end{center}
\end{table}

Including the isospin-breaking corrections, one obtains the results given in Table~\ref{tab:ib-lo-nlo}. 
The  primary fitted quantities $\text{Re}\, g_8$, $\text{Re}\,g_{27}$ and $\chi_0-\chi_2$,  as well as the derived quantities (such as  $\text{Re}\,A_{0,2}$), 
depend now on the adopted $\chi$PT approximation, LO or NLO.
The experimental uncertainties are again indicated by the first errors. 
Moreover, the presence of an $\cO(e^2p^0)$ electromagnetic-penguin contribution makes the LO fit also sensitive to the short-distance scale $\mu_{\rm SD}$ (second errors). Our LO results are in agreement with the Flavianet averages \cite{Antonelli:2010yf} in Eq.~\eqn{eq:isoamps}. At the NLO, the presence of the electromagnetic correction $f_{5/2}$ implies that $\text{Re}\,A_2^+ \not= \text{Re}\,A_2$. The NLO results have explicit dependencies on both renormalization scales, $\nu_\chi$ (second errors) and $\mu_{\rm SD}$ (third errors). Notice that the isotensor amplitude and $g_{27}$ are quite sensitive to the isospin-breaking corrections.

\begin{table}[th]
\begin{center}
\begin{tabular}{|c|c|c|}\hline
     & LO fit & NLO fit \\ \hline
$\real \:g_8$ & $5.002 \pm 0.002_{\,\rm exp}\, {^{\,+\,0.008}_{\,-\,0.004}}_{\,\mu_{\rm SD}}$ 
& $3.582 \pm 0.001_{\,\rm exp} \, {^{\,+\,0.144}_{\,-\,0.141}}_{\,\nu_\chi}\, {^{\,+\,0.016}_{\,-\,0.006}}_{\,\mu_{\rm SD}}$
\\\hline
$\real \:g_{27}$ & $0.251 \pm  0.001_{\,\rm exp}  \, {^{\,+\,0.007}_{\,-\,0.003}}_{\,\mu_{\rm SD}}$ 
& $0.297 \pm 0.001_{\,\rm exp} \, {^{\,+\,0.000}_{\,-\,0.001}}_{\,\nu_\chi}\, {^{\,+\,0.006}_{\,-\,0.002}}_{\,\mu_{\rm SD}} $
\\\hline
 $\chi_0-\chi_2\; (^{\circ})$ & $47.97\pm  0.92_{\,\rm exp}  \, {^{\,+\,0.08}_{\,-\,0.16}}_{\,\mu_{\rm SD}}$
 & $51.396 \pm 0.806_{\,\rm exp} \, {^{\,+\,1.041}_{\,-\,1.051}}_{\,\nu_\chi} \, {^{\,+\,0.017}_{\,-\,0.003}}_{\,\mu_{\rm SD}}$
 \\\hline
 \hline
 $\real \:A_0\; (10^{-7}\:\text{GeV})$ & $2.704 \pm  0.001_{\,\rm exp}$  
& $2.704 \pm 0.001_{\,\rm exp}$
\\\hline
$\real \:A_2\; (10^{-8}\:\text{GeV})$ & $1.222 \pm  0.003_{\,\rm exp}\, {^{\,+\,0.002}_{\,-\,0.004}}_{\,\mu_{\rm SD}}$ 
& $1.317 \pm 0.003_{\,\rm exp}  \, {^{\,+\,0.033}_{\,-\,0.031}}_{\,\nu_\chi} \, {^{\,+\,0.001}_{\,-\,0.000}}_{\,\mu_{\rm SD}}$
\\\hline
  $f_{5/2}$ & 0 
 & $0.0852 \pm 0.0002_{\,\rm exp} \, {^{\,+\,0.0239}_{\,-\,0.0250}}_{\,\nu_\chi} \, {^{\,+\,0.0001}_{\,-\,0.0004}}_{\,\mu_{\rm SD}}$
 \\\hline
 $\real \: A_0/\real \: A_2$  & $22.13 \pm  0.05_{\,\rm exp} \, {^{\,+\,0.07}_{\,-\,0.04}}_{\,\mu_{\rm SD}}$
 & $20.54\pm 0.04_{\,\rm exp} \, {^{\,+\,0.50}_{\,-\,0.50}}_{\,\nu_\chi} \, {^{\,+\,0.00}_{\,-\,0.01}}_{\,\mu_{\rm SD}}$ 
 \\\hline
 $\real \:A_0/\real \:A_2^+$& $22.13 \pm  0.05_{\,\rm exp} \, {^{\,+\,0.07}_{\,-\,0.04}}_{\,\mu_{\rm SD}}$
& $22.28 \pm 0.05_{\,\rm exp} \, {^{\,+\,0.01}_{\,-\,0.06}}_{\,\nu_\chi} \, {^{\,+\,0.00}_{\,-\,0.02}}_{\,\mu_{\rm SD}}$ 
 \\\hline
\end{tabular}
\caption{\label{tab:ib-lo-nlo} LO and NLO fits to the $K\to\pi\pi$ amplitudes, including isospin breaking.}
\end{center}
\end{table}

The results in Tables \ref{tab:ic-lo-nlo} and \ref{tab:ib-lo-nlo}  supersede 
the values obtained in Ref.~\cite{Cirigliano:2003gt}. The main differences originate in the more precise experimental data now available.

\subsection{Isospin-breaking parameters in the CP-odd sector}

We have now all the needed ingredients to compute the different isospin-breaking (IB) parameters in the CP-odd sector, defined in Section~\ref{sec:Decomp}. The resulting values are displayed in Table~\ref{tab:IB-results} at different levels of approximation. The first two columns show the results obtained with $\alpha=0$ at LO and NLO, respectively; {\it i.e.} they refer to strong isospin violation only ($m_u \not = m_d$). The impact of electromagnetic corrections is shown in the last two columns, which contain the complete results including electromagnetic corrections.

\begin{table}[th]
\centering
\begin{tabular}{|c|cc|cc|}
\hline
& \multicolumn{2}{|c|}{$\alpha =0$} & \multicolumn{2}{|c|}{$\alpha\not= 0$}
\\
& LO & NLO  & LO & NLO
\\ \hline
$\Omega_{\text{IB}}$ & $13.7$  & $15.9\pm 8.2$ & $19.5\pm 3.9$ & $24.7\pm 7.8$
\\
$\Delta_{0}$ & $-0.002$ & $-0.49\pm 0.13$ & $5.6 \pm 0.9$ & $5.6\pm 0.9$
\\
$f_{5/2}$ & 0 & 0 & 0 & $8.2\, ^{\,+\,2.3}_{\,-\,2.5}$
\\ \hline
$\Omega_{\text{eff}}$ & $13.7$ & $16.4\pm 8.3$ & $13.9\pm 3.7$ & $11.0\, {^{\,+\,9.0}_{\,-\,8.8}}$
\\ \hline
\end{tabular}
\caption{Isospin-violating corrections for $\epsilon'/\epsilon$ in units of $10^{-2}$. }
\label{tab:IB-results}
\end{table}


In Appendix~\ref{app:results} we provide a detailed comparison with the results of Ref.~\cite{Cirigliano:2003nn,Cirigliano:2003gt}, analyzing the impact of the different updated inputs in the final NLO values. The most significant changes are a slight reduction of the IB correction to $A_0$, $\delta\Delta_0 \approx -0.028$, induced by the numerical changes in $L_5$ and the Wilson coefficients, and an increased value of $\Omega_{\rm IB}$, $\delta\Omega_{\rm IB}\approx 0.020$, which is mostly driven by $L_7$ (there are also sizeable changes from $L_5$, $K_i$ and $\varepsilon^{(2)}$ that cancel among them to a large extent). The net combined effect is a larger central value of the global correction $\delta\Omega_{\text{eff}}\approx 0.05$.
The largest sources of uncertainty turn out to be the input values of the strong LECs $L_7$, $L_5$ and $L_8$ (parametric)  and the dependence on 
the chiral renormalization scale $\nu_\chi$ (a ``systematic error"  induced by the large-$N_C$ approximation).
Appendix~\ref{app:results} contains a detailed description of the different errors.

\begin{figure}[h!]
\begin{center}
\includegraphics[scale=0.7]{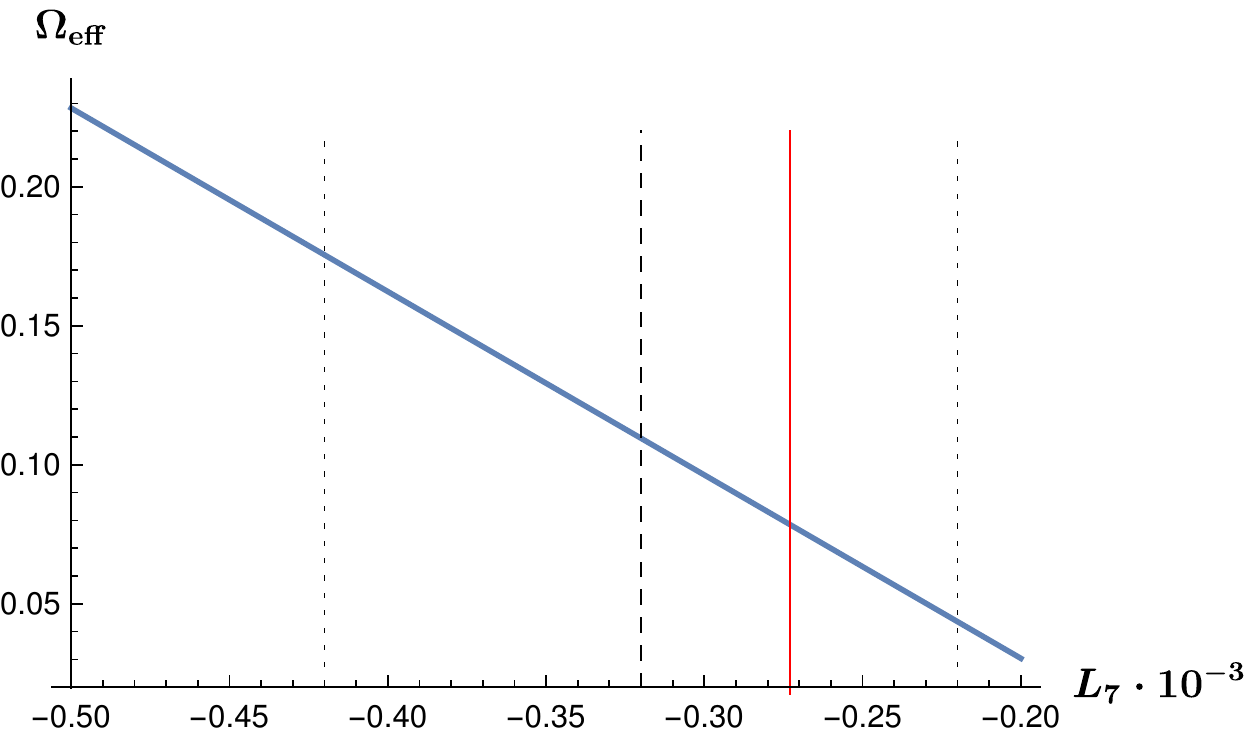}
\end{center}
\caption{Central value of $\Omega_{\text{eff}}$ as a function of $L_7$.
The dotted vertical lines indicate the  range of $L_7$ in Eq.~\eqn{eq:Li-inputs}, while the red line is the large-$N_C$ value from Eq.~\eqn{eq:L7_LargeNc}. 
}\label{fig:omegaeffL7}
\end{figure}

The final prediction for $\Omega_{\text{eff}}$ is very sensitive to the input value of $L_{7}$. Figure~\ref{fig:omegaeffL7} illustrates the strong dependence of the central value of $\Omega_{\text{eff}}$ with $L_7$. The dashed vertical line indicates the  value of $L_7$ in Eq.~\eqn{eq:Li-inputs}  \cite{Bijnens:2014lea}, with its error range (dotted lines). The red line is the large-$N_C$ prediction for $L_7$ in Eq.~\eqn{eq:L7_LargeNc}. 

We conclude this section by discussing the applicability of our 
results on isospin-breaking effects in $\epsilon'$,   obtained in the framework of $\chi$PT, 
 to other non-perturbative methods,  that typically estimate hadronic matrix elements 
in the isospin limit (see for example Refs.~\cite{Bai:2015nea,Buras:2015yba}).  
Our  two main observations are:

\begin{itemize}

\item First,   $\Delta_0$ is largely dominated by electromagnetic 
penguin contributions. Therefore,  in those theoretical calculations of $\epsp$
where electromagnetic penguin contributions are explicitly included 
in  $A_0$,  one should  remove their effect  from  the quantity $\Delta_0$, 
keeping only the strong isospin-breaking contributions to this quantity.
This amounts to the replacement   $ \Omega_{\rm eff} \to  \hat{\Omega}_{\text{eff}}$ 
with~\cite{Buras:2015yba,Cirigliano:2003gt}
\beq
     \hat{\Omega}_{\text{eff}}\:\equiv\:\Omega_{\text{IB}}\:-\:\Delta_0|_{\alpha=0}\:-\:f_{5/2}~,
\eeq
since $\Delta_0$ is the only contribution proportional to $\IM A_0$. The updated value is
 \begin{align} 
     \hat{\Omega}_{\text{eff}}\:=\:(17.0\:{}^{+9.1}_{-9.0})\cdot 10^{-2}~,  
 \end{align}
which can be directly extracted from Table~\ref{tab:IB-results}. The final error has been obtained taking into account the correlation among those values.

\item  Second,  in applying  isospin-breaking corrections 
one needs to keep track of how  isospin-symmetric QCD is defined in each calculation. 
This intrinsically implies a scheme dependence  (see \cite{Aoki:2019cca,DiCarlo:2019thl} and references therein).
In Appendix~\ref{app:scheme} we have presented the separation scheme adopted in this work (following~\cite{Cirigliano:2003gt}) 
and a possible alternative scheme.  We have then discussed the implications of scheme dependence for 
$\Omega_{\rm eff}$,  finding that, for the two schemes considered, 
the numerical effect is  well below  current  theoretical uncertainties. 

\end{itemize}

\section{Updated SM prediction for $\boldsymbol{\epsilon'/\epsilon}$}
\label{sect:conclusions}

\begin{figure}[h]
\begin{center}
\includegraphics[scale=0.7]{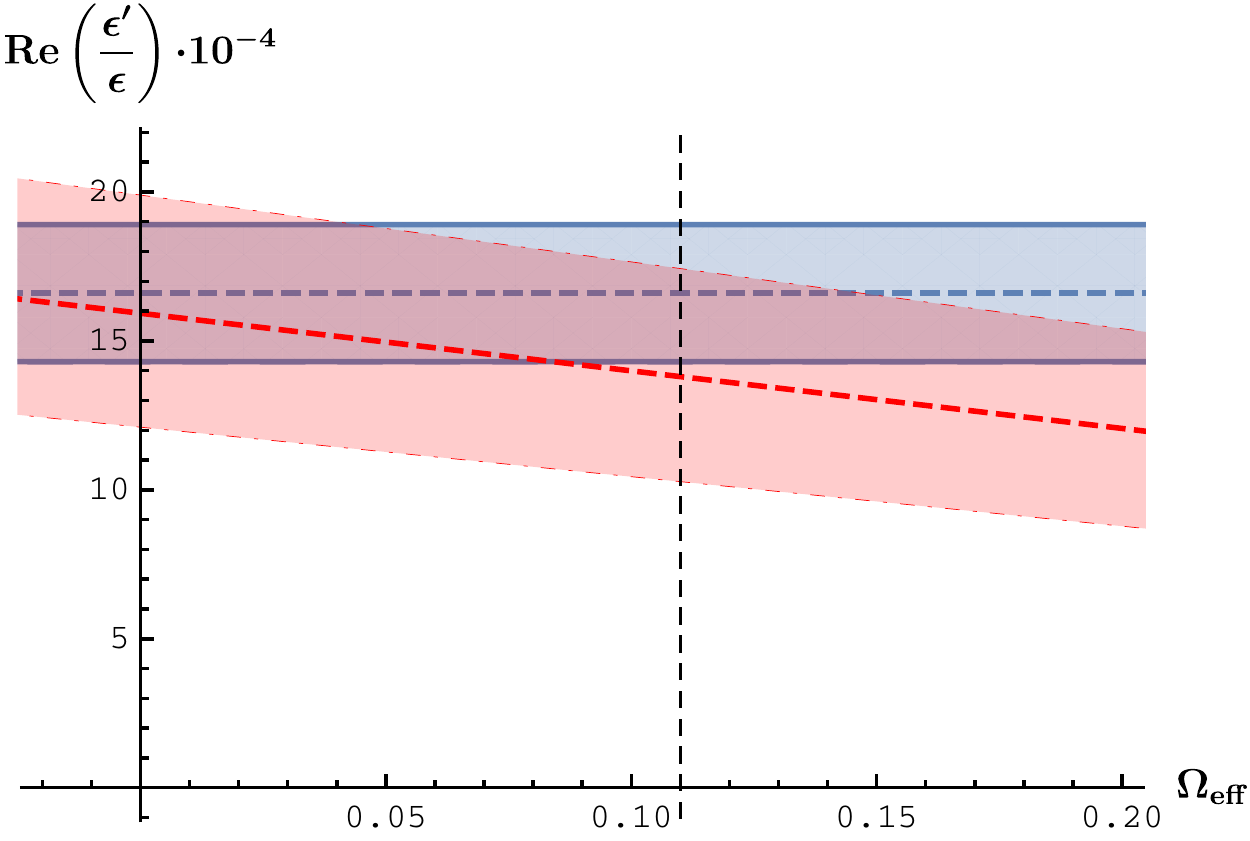}
\end{center}
\caption{SM prediction for $\text{Re}\left(\epsilon'/\epsilon\right)$ (red dashed line) as a function of $\Omega_\text{eff}$. The red band has been obtained adding all sources of uncertainty 
in quadrature for a fixed value of $\Omega_{\rm eff}$. The vertical dashed line indicates the central value of $\Omega_\text{eff}$ in \eqn{eq:Omega_eff_res} and the blue horizontal band the measured value of $\text{Re}\left(\epsilon'/\epsilon\right)$.}\label{fig:omegaeff}
\end{figure}

The improved knowledge on many of the inputs entering the calculation of isospin-breaking corrections to the $K\to\pi\pi$ amplitudes has allowed us to perform a thorough numerical update of the pioneering analysis of Ref.~\cite{Cirigliano:2003nn,Cirigliano:2003gt}. We have presented in this paper a comprehensive review of the theoretical approach and have discussed in detail the different parametric improvements and their impact on the relevant isospin-breaking contributions. Our final result for the key parameter in the CP-odd sector is (see 
Eqs.~(\ref{eq:cpeff}), (\ref{eq:omegaeff})  and Table~\ref{tab:IB-results}):
\begin{align}\label{eq:Omega_eff_res}
\Omega_{\text{eff}}\:=\:(11.0^{\,+\,9.0}_{\,-\,8.8})\cdot 10^{-2}\, ,
\end{align}
where the final uncertainty has been obtained by conservatively adding all errors in quadrature.

Figure~\ref{fig:omegaeff} shows the dependence of $\text{Re}\left(\epsilon'/\epsilon\right)$ on $\Omega_{\text{eff}}$. Taking into account the updated value of this parameter, our SM prediction for $\text{Re}\left(\epsilon'/\epsilon\right)$, 
\begin{align}\label{eq:epsp_prediction}
\text{Re}\left(\epsilon'/\epsilon\right)\,&=\,\left( 13.8{^{\,+\,0.5}_{\,-\,0.4}}_{\,m_s} {^{\,+\,1.7}_{\,-\,1.3}}_{\,\mu_{\rm SD}} {^{\,+\,3.1}_{\,-\,3.2}}_{\,\nu_\chi}\pm 1.3_{\,\gamma_5} \pm 2.1_{\,L_{5,8}} \pm 1.3_{\,L_7} \pm 0.2_{\,K_i} \pm 0.3_{\,X_i}\right)\cdot 10^{-4}
\nonumber\\[0.25cm]
&=\:\left(14\,\pm\,5\right)\cdot 10^{-4} \, ,
\end{align}
is in excellent agreement with the experimental world average \cite{Batley:2002gn,Lai:2001ki,Fanti:1999nm,Barr:1993rx,Burkhardt:1988yh,Abouzaid:2010ny,AlaviHarati:2002ye,AlaviHarati:1999xp,Gibbons:1993zq},
\begin{align}
\text{Re}\left(\epsilon'/\epsilon\right)_{\text{exp}}\:&=\:\left(16.6 \pm 2.3\right)\cdot 10^{-4}\, .
\end{align}

In Eq.~\eqref{eq:epsp_prediction}, we display the different sources of uncertainty in $\text{Re}\left(\epsilon'/\epsilon\right)$. The first error represents the sensitivity to the input quark masses. Our ignorance about $1/N_{C}$-suppressed contributions in the matching region is parametrized through the second and third errors, which have been estimated very conservatively through the variation of $\mu_{\rm SD}$ and $\nu_{\chi}$ in the intervals $ [0.9,1.2] \, \mathrm{GeV}$ and $[0.6,1]$~GeV, respectively. The fourth error reflects the choice of scheme for $\gamma_5$. The fifth and sixth errors originate from the input values of the strong LECs $L_{5,7,8}$, given by Eq.~\eqref{table:L578}, and the last two errors correspond to the uncertainties of the NLO electromagnetic LECs $K_{i}$ and the NNLO strong couplings $X_{i}$; they have been estimated using Eq.~\eqref{eq:XiError}.

\begin{figure}[h]
\begin{center}
\includegraphics[scale=0.8]{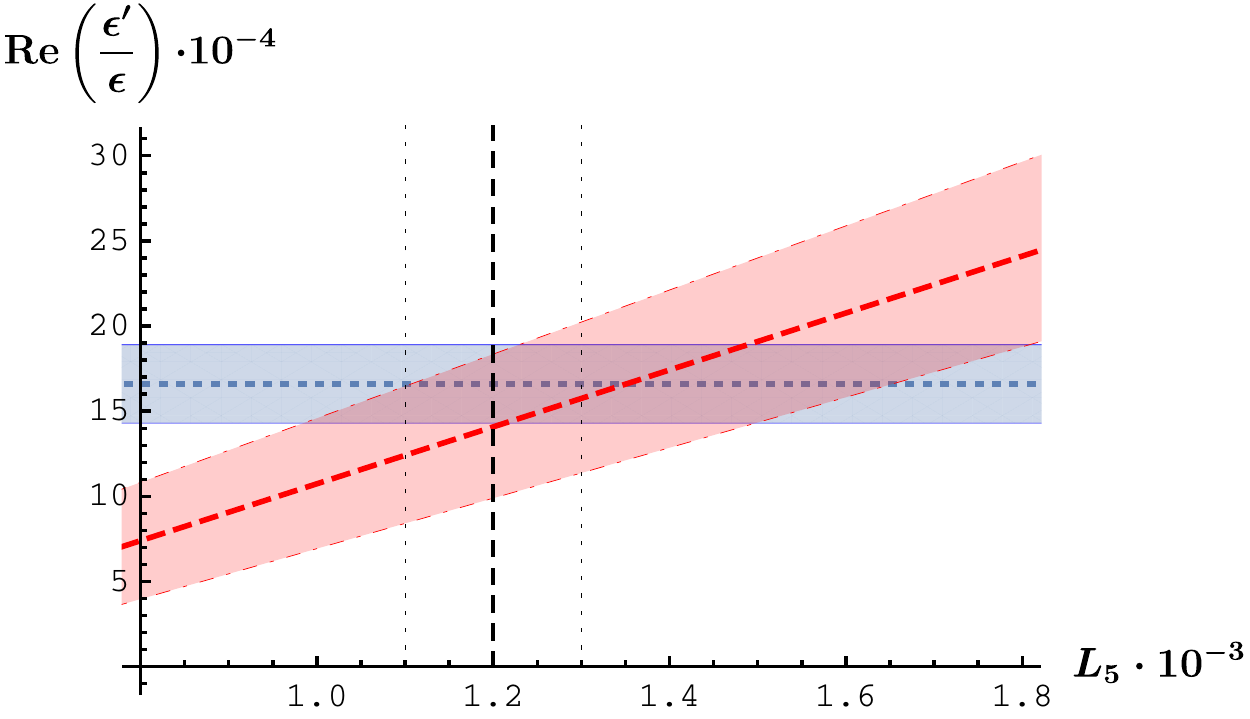}
\end{center}
\caption{
SM prediction for $\text{Re}\left(\epsilon'/\epsilon\right)$ (red dashed line) as a function of $L_5$. The value of $L_8$ has been fixed in terms of $L_5$, using their relation in Eq.~\eqref{eq:Li-inputs}. The red band has been obtained adding all sources of uncertainty 
in quadrature for a fixed value of $L_5$. The black dashed vertical lines represent the central value of $L_5^r(M_\rho)$ with its error, given in Eq.~\eqref{eq:Li-inputs}. The blue horizontal band is the measured value of $\text{Re}\left(\epsilon'/\epsilon\right)$.}\label{fig:L5}
\end{figure}

The updated value of $\Omega_{\text{eff}}$  has a relatively small numerical impact on the final prediction for $\epsilon'/\epsilon$, giving a central value slightly smaller than the one obtained in Ref.~\cite{Gisbert:2017vvj} with the old IB inputs. The large theoretical uncertainty in \eqn{eq:epsp_prediction}, mostly coming from our ignorance of non-perturbative effects in the matching region and the strong dependence on the parameter $L_5$ (see Figure~\ref{fig:L5}), has been estimated conservatively and could be reduced in the future. A detailed discussion of other possible improvements was presented in Ref.~\cite{Gisbert:2017vvj}.

\section*{Acknowledgements}
We warmly acknowledge early collaboration with Gerhard Ecker and Helmut Neufeld, and useful discussions with Hans Bijnens, Gerhard Ecker and Chris Sachrajda. 
This work has been supported in part by the Spanish Government and ERDF funds from
the EU Commission [grant FPA2017-84445-P], the Generalitat Valenciana [grant Prometeo/2017/053], the Spanish Centro de Excelencia Severo Ochoa Programme [grant SEV-2014-0398],
the Swedish Research Council [grants 2015-04089  and  2016-05996]  and  by  the  European  Research Council (ERC) under the EU Horizon 2020 research and innovation programme (grant 668679). The work of H.G. is supported by a FPI doctoral contract [BES-2015-073138], funded by the Spanish Ministry of Economy, Industry and Competitiveness and the Bundesministerium f\"ur Bildung und Forschung (BMBF).
V.C. acknowledges support by the US DOE Office of Nuclear Physics.

\appendix

\section{Parameters of large-$\boldsymbol{N_C}$ matching at NLO}
\label{sec:largeNCparameters}

\begin{table}[h!]
\begin{center}
\begin{tabular}{|c|c|c|}
\hline
$i$ & $n_i$ & $\mathcal{X}_i$    \\ \hline
5  & -2& $ - 16 \,  X_{14}
+ 32 \,  X_{17}
- 24 \,  X_{38}
- 4 \,  X_{91}$ \\ \hline
6  &  4& $- 32 \,  X_{17}
- 32 \,  X_{18}
+ 32  \,  X_{37}
+ 16  \,  X_{38}$  \\ \hline
7  &  2&  $- 32 \,  X_{16}
- 16 \,  X_{17}
+ 8  \,  X_{38}$  \\ \hline
8  &  4& $- 16 \,  X_{15}
- 32 \,  X_{17}
+ 16 \,  X_{38}$  \\ \hline
9  &  0&  $- 64 \, L_5 \,  L_8 
- 8 \,  X_{34}
+ 8  \,  X_{38}
+ 4 \,  X_{91}$ \\ \hline
10 &  0&  $- 48 \,  X_{19}
- 8 \,  X_{38}
- 2 \,  X_{91}
- 4 \,  X_{94}$ \\ \hline
11 &  0& $- 32 \,  X_{20}
+ 4 \,  X_{94}$  \\ \hline
12 &  0& $128 \, L_8 \, L_8  
+ 16 \,  X_{12}
- 16 \,  X_{31}
+ 8 \,  X_{38}
- 2 \,  X_{91}
- 4 \,  X_{94}$  \\ \hline
13 &  0& $256  \, L_7 \,  L_8 
- \frac{32}{3} \,  X_{12}
- 16 \,  X_{33}
+ 16 \,  X_{37}
+ \frac{4}{3} \,  X_{91}
+ 4 \,  X_{94}$  \\ \hline
\end{tabular}
\end{center}
\caption{Parameters $n_i$ and $\mathcal{X}_i$ entering the prediction
of the LECs $(g_8\:N_i)^\infty$ in Eq.~\eqn{eq:g8Ni}.}
\label{table:g8N_i}
\end{table}

Table~\ref{table:g8N_i} compiles the values of $n_i$ and $\mathcal{X}_i$ that parametrize the 
large-$N_C$ predictions for the weak LECs $(g_8\:N_i)^\infty$ in Eq.~\eqn{eq:g8Ni}. The $\mathcal{X}_i$ parameters are functions of the strong $\cO(p^6)$ couplings $X_i$.
The LEC $X_{94}$ only appears in $\mathcal{X}_i$ for $i=10,11,12,13$. The corresponding couplings $N_i$ contribute to $\Delta_C\cA^{(8)}_{1/2}$ and $\Delta_C\cA^{(\varepsilon)}_{1/2,3/2}$, but always in combinations of the form $\sum_{i=10}^{13} a_i N_i$ with $a_{10}+a_{12}=a_{11}+a_{13}$.
Thus, $X_{94}$ drops completely from the $K\to\pi\pi$ amplitudes. The same happens with $X_{37}$, because $\mathcal{X}_6$ and $\mathcal{X}_{13}$ only enter through the combination $N_6^r-2 N_{13}^r$.

The large-$N_C$ predictions for the $\cO(p^6)$ LECs $X_i$ were estimated in Ref.~\cite{Cirigliano:2006hb} through resonance exchange. The role of the $\eta_1$ meson in these LECs was further analyzed in Ref.~\cite{Kaiser:2007zz}. The only $\eta_1$-exchange contributions
to the $K\to\pi\pi$ amplitudes are
\begin{align}
\widetilde{X}^{\eta_1}_{18}\:=\:3\:\widetilde{X}^{\eta_1}_{19}\:=\:-\:2\:\widetilde{ X}^{\eta_1}_{20}\:=\:\widetilde{X}^{\eta_1}_{31}\:=\:\frac{L_7^\infty}{M_{\eta_1}^2}\, ,
\qquad \qquad
\widetilde{X}^{\eta_1}_{33}\:=\:0 \, .
\end{align}

\begin{table}[h!]
\begin{center}
\resizebox{1.0\textwidth}{!}{
\begin{tabular}{|c|c|c|c|c|c|c|c|}
\hline
$i$ & $\mathcal{K}^{(1)}_i$ & $\mathcal{K}^{(2)}_i$ & $\mathcal{K}^{(3)}_i$ & $\mathcal{K}^{(4)}_i$ & $\mathcal{K}^{(5)}_i$ & $\mathcal{K}^{(6)}_i$ &  $\mathcal{K}^{(7)}_i$  \\ \hline
1  & $\frac{1}{3} \, K_{12} -  K_{13}$ & 0 & $ 64\, L_{8} \, ( 
-\frac{1}{3} \, K_{9}  + \frac{5}{3} \, K_{10}  + K_{11}   
)$ & 0 & $- 24\, L_8$ & 0 & 0   \\ \hline
2  & $\frac{4}{3}\, K_{13}$& 0 & $- \frac{256}{3}\, L_8\, 
\left( K_{10} + K_{11} \right)$ & 0 & 0 & 0 & 0   \\ \hline
3  & $K_{13}$ & 0 & $- 64 \, L_8\, 
\left( K_{10} + K_{11} \right) $& 0 & 0 & 0 & 0  \\ \hline
4  & $-K_{13}$& 0 & $64 \, L_8\, 
\left( K_{10} + K_{11} \right) $ & 0 & 0 & 0 & 0  \\ \hline
5  & $\frac{4}{3} \, 
\left( 4 \, K_1 + 3 \, K_{5} + 3 \, K_{12} 
\right) $& 0 & $- \frac{64}{3}\, L_5 \, 
\left( 2 K_{7} + K_{9} \right)   $ & 0 & 0 & 0 & $1$  \\ \hline
6  & $ -\frac{2}{3} \,  ( K_{5} + K_{6} )  + 2  \, (K_{12} + K_{13}) $ & 0 & $- \frac{32}{3}\, L_5 \,  
\left( K_{9} + K_{10} + 3 K_{11} \right)  $ & 0 & $-12\, L_5$ & 0 & 0  \\ \hline
7  & $8 \, K_2 + 6 \, K_{6}  - 4 \, K_{13}$ & 0 & $- 32 \, L_5\,   
\left(2 K_{8} + K_{10} + K_{11} \right)   $ & 0 & 0 & 0 & 0  \\ \hline
8  & $\frac{8}{3} \, K_{3} + 4 \, K_{12} $ & $\frac{4}{3}\, K_5$ & 0 & 0 & 0 & $\frac{3}{2}$ & $\frac{3}{2}$  \\ \hline
9  & $- \frac{4}{3} \, \left( K_4 + K_{12} + 
K_{13} \right)$& $\frac{4}{3}\, K_5$ & 0 & $-\frac{3}{2}$ & 0 & 0 & 0   \\ \hline
10 & $- 2  \, K_{13}$& $4\, K_6$& 0 & 0 & 0 & 0 & 0  \\ \hline
11 & $2 \,  \left( K_4 + K_{13} \right) $& 0 & 0 & 0 & 0 & 0 & 0  \\ \hline
12 & $- 4 \, K_3 $ &  0 & 0 & 0 & 0 & 0 &  0 \\ \hline
\end{tabular}}
\end{center}
\caption{Large-$N_C$ parameters $\mathcal{K}^{(k)}_i$ of the $(g_8\, Z_i)^\infty$ LECs in Eq.~\eqn{eq:g8Zi-LargeNc}.}
\label{table:g8Z_i}
\end{table}

The large-$N_C$ predictions for the electroweak LECs $(g_8\, Z_i)^\infty$ in Eq.~\eqn{eq:g8Zi-LargeNc}
are governed by the constants $\mathcal{K}^{(k)}_i$, compiled in Table~\ref{table:g8Z_i}.
They are functions of the electromagnetic and strong $\chi$PT couplings $K_i$ and $L_i$, respectively.

\section{Updated estimate of $\boldsymbol{\lambda_3^{SS}}$}
\label{app:lambda3}

The R$\chi$T coupling $\lambda_3^{SS}$ splits the masses of the different isospin components of the scalar-resonance nonet multiplet through the term
\begin{equation}
\mathcal{L}_{\text{S}}^{\mathrm{mass}}\, =\, - \frac{M_S^2}{2}\; \langle S^2\rangle
+ \lambda_3^{SS}\; 4 B_0\; \langle S^2\cM\rangle\, .
\end{equation}
The common multiplet mass and  $\lambda_3^{SS}$ can then be determined through the relations~\cite{Cirigliano:2003yq}:
\begin{equation}
\lambda_3^{SS}\:=\:\frac{M_{I=1}^2\:-\:M_{I=1/2}^2}{4\:(M_K^2\:-\:M_\pi^2)}\, ,
\qquad\qquad 
M_S^2\:=\: M_{I=1}^2\:+\:\frac{M_\pi^2\:(M_{I=1}^2\:-\:M_{I=1/2}^2)}{M_K^2\:-\:M_\pi^2}\, ,
\label{eq:masscoupling}   
\end{equation}
with $M_I$ the mass of the scalar meson with isospin $I$.

In order to identify the members of the scalar resonance nonet, we must exclude the lightest observed scalars that are well understood as dynamically-generated poles arising from 2-Goldstone scattering:
$f_0(500)$ ($\sigma$), $K_0^*(700)$ ($\kappa$), $a_{0}(980)$ and $f_0(980)$ \cite{Oller:2003vf,Jamin:2000wn,Ledwig:2014cla,Caprini:2005zr,Pelaez:2015qba}.
The $I=1/2$ and $I=1$  members of the resonance nonet are identified without controversy with $K^{*}_{0}(1430)$  and $a_0(1450)$ respectively. For the $I=0$ states, we have three possible candidates: $f_0(1370)$, $f_0(1500)$ and $f_0(1710)$. Thus, there are two possible scenarios:
\begin{align*}
A:\qquad f_0(1370),\:\:K^*_0(1430),\:\:a_0(1450),\:\:f_0(1500).\\[0.5cm]
B:\qquad f_0(1370),\:\:K^*_0(1430),\:\:a_0(1450),\:\:f_0(1710).
\end{align*}
One can figure out the favoured dynamical option, comparing these candidates with the predicted isosinglet masses. Using the relation~\cite{Cirigliano:2003yq},
\begin{equation}
M_{L,H}^2\:=\:M_{I=1/2}^2\:\mp\:|M_{I=1/2}^2\:-\:M_{I=1}^2|\, ,
\end{equation}
we find  $M_L=1374$ MeV and $M_H=1474$ MeV for the lighter and heavier isosinglet scalar states, respectively.
Therefore, we can conclude that the lightest scalar-resonance nonet is given by the scenario A. Moreover, since the values of $M_{L,H}$ are very close to the measured masses, additional nonet-symmetry-breaking corrections to the scalar masses can be neglected ({\it i.e.}, $k_m^R=\gamma_R=0$, in Ref.~\cite{Cirigliano:2003yq}). Inserting the scalar resonance masses in the relations~\eqn{eq:masscoupling}, one finally finds the values of $M_S$ and $\lambda_3^{SS}$ given in Eq.~\eqn{eq:massscalar}.

\section{Parametric uncertainties in $\boldsymbol{\Omega_{\textbf{eff}}}$,  $\boldsymbol{\Omega_{\textbf{IB}}}$, $\boldsymbol{\Delta_0}$ and  $\boldsymbol{f_{5/2}}$}
\label{app:results}

Since this work is an update of Ref.~\cite{Cirigliano:2003nn,Cirigliano:2003gt}, it is worth to compare the impact of the different updated inputs in the final (central) values of the IB parameters.
This is shown in Table~\ref{tab:updcomp} for the results of the complete NLO analysis with $\alpha\neq 0$. The quantities $\Delta_{i}$ correspond to the difference between the updated result and the one obtained with the old input for the variable $i$ ($i={\rm WC}$ stands for Wilson Coefficients). The impact of the different changed inputs is comparable in size, and typically slightly smaller than the central values. In particular, the sensitivity to $L_{7}$ is remarkable.

\begin{table}[htb!]
\centering
\begin{tabular}{|c|c|c|c|c|}
 \hline
Set-up & $\Delta_{0}$ & $f_{5/2}$ & $\Omega_{\text{IB}}$ & $\Omega_{\text{eff}}$ \\ \hline
Old value \cite{Cirigliano:2003gt} & $0.08346$ & $0.08360$ & $0.2267$ & $0.05967$\\ \hline
New value & $0.05578$ & $0.08168$ & $0.2470$ &  $0.1095$\\ \hline
$\Delta_{\rm WC}$ & $-0.11$ & $-0.0008$ & $0.0017$ & $0.013$\\ \hline
$\Delta_{L_{5}}$ & $-0.017$ & $0.0009$ & $-0.032$ & $-0.016$   \\ \hline
$\Delta_{L_{8}}$ & $0.0028$ & $0.0012$ & $-0.0060$ & $-0.010$ \\ \hline
$\Delta_{L_{7}}$ & $-0.0006$ & $0.0000$ & $0.029$ & $0.029$ \\ \hline
$\Delta_{K_{i}}$ & $0.0012$ & $-0.0036$ & $0.022$ & $0.024$\\ \hline
$\Delta_{X_{i}}$ & $0.0017$ & $0.0001$ & $-0.0011$ & $-0.0029$\\ 
\hline
$\Delta_{\varepsilon^{(2)}}$ & $-0.0003$ & $0.0000$  & $0.011$  & $0.011$  \\ 
\hline
$\Delta_{B(\mu_{\rm SD})}$ & $-0.0049$  &  $0.0005$ &  $-0.0066$ & $-0.0021$  \\ 
\hline
\end{tabular}
\caption{\label{tab:updcomp}NLO central values for $\alpha\neq 0$ and impact of the different modified inputs.}
\end{table}

In Tables \ref{tabnlo0}, \ref{tablo} and \ref{tabnlo} we detail the different sources of parametric uncertainties 
for $\Delta_0$, $f_{5/2}$, $\Omega_{\text{IB}}$, and $\Omega_{\text{eff}}$ at both LO and NLO,  and for $\alpha=0$ and $\alpha \neq0$.
We consider the following uncertainties:
\begin{itemize}
    \item $\sigma_{\mu_{\rm SD}}$ and $\sigma_{\nu_{\chi}}$. Uncertainties associated to the large-$N_{C}$ matching procedure, which leads to ambiguities when setting both the short-distance ($\mu_{\rm SD}$) and the chiral ($\nu_{\chi}$) scales. They are estimated by varying them in the intervals $\mu_{\rm SD}\in [0.9,1.2] \, \mathrm{GeV}$ and $\nu_{\chi}\in [0.6,1]$~GeV.
    \item  $\sigma_{\gamma_5}$. Uncertainty associated with the choice of renormalization prescription for $\gamma_{5}$. We have taken the difference between the results obtained using the HV and NDR schemes.
    \item $\sigma_{L_{5,7,8}}$. Uncertainties from the input values of the strong LECs $L_{5,7,8}$ in Eq. (\ref{table:L578}).
   \item $\sigma_{K_{i}}$ and $\sigma_{X_{i}}$. Uncertainties associated, respectively, with the NLO electromagnetic LECs $K_{i}$ and the NNLO strong couplings $X_{i}$.
\end{itemize}

\begin{table}[htb!]
\centering
\begin{tabular}{|c|c|c|c|c|}
 \hline
Set-up & $\Delta_{0}$ & $f_{5/2}$ & $\Omega_{\text{IB}}$  & $\Omega_{\text{eff}}$ \\ \hline
Central& $-0.0049$ & $0.0$ & $0.159$ & $0.164$ \\ \hline
$\sigma_{\mu_{\rm SD}}$& $^{\,+\,0.0001}_{\,-\,0.0002}$ & $0.0$ & $^{\,+\,0.001}_{\,-\,0.001}$& $^{\,+\,0.001}_{\,-\, 0.001}$ \\ \hline
$\sigma_{\nu_\chi}$& $0.0001$ & $0.0$ & $^{\,+\,0.048}_{\,-\,0.047}$ &  $^{\,+\,0.048}_{\,-\,0.047}$ \\ \hline
$\sigma_{\gamma_5}$ & $0.0004$ & $0.0$ & $0.001$ & $0.002$ \\ \hline
$\sigma_{L_{5,8}}$ & $0.0001$ & $0.0$ & $0.015$ & $0.015$ \\ \hline
$\sigma_{L_{7}}$ & $0.0012$ & $0.0000$ & $0.065$ &  $0.066$ \\ \hline
$\sigma_{X_{i}}$ & $0.0000$ & $0.0$ & $0.007$ & $0.007$\\ 
\hline
\end{tabular}
\caption{\label{tabnlo0} NLO central values for $\alpha=0$ and their parametric errors.}
\end{table}

\begin{table}[htb!]
\centering
\begin{tabular}{|c|c|c|c|c|}
 \hline
Set-up & $\Delta_{0}$ & $f_{5/2}$ & $\Omega_{\text{IB}}$ &  $\Omega_{\text{eff}}$ \\ \hline
Central& $0.0557$ & $0.0$ & $0.195$ & $0.139$ \\ \hline
$\sigma_{\mu_{\rm SD}}$&  $^{\,+\,0.0003}_{\,-\,0.0000}$ & $0.0$ & $^{\,+\,0.001}_{\,-\,0.001}$ & $^{\,+\,0.001}_{\,-\,0.001}$\\ \hline
$\sigma_{\nu_\chi}$& $0.0000$ & $0.0$ & $0.000$ & $0.000$\\ \hline
$\sigma_{\gamma_5}$ & $0.0066$ & $0.0$ & $0.001$ & $0.006$\\ \hline
$\sigma_{L_{5,8}}$ & $0.0053$ & $0.0$ & $0.010$ & $0.005$ \\ \hline
$\sigma_{K_{i}}$& $0.0021$ & $0.0$ & $0.038$ & $0.036$ \\ 
\hline
\end{tabular}
\caption{\label{tablo} LO central values for $\alpha\neq 0$ and their parametric errors. }
\end{table}

\begin{table}[htb!]
\centering
\begin{tabular}{|c|c|c|c|c|}
 \hline
Set-up & $\Delta_{0}$ & $f_{5/2}$ & $\Omega_{\text{IB}}$ &  $\Omega_{\text{eff}}$ \\ \hline
Central& $0.0558$ & $0.0817$ & $0.247$ &  $0.110$ \\ \hline
$\sigma_{\mu_{\rm SD}}$&  $^{\,+\,0.0014}_{\,-\,0.0011}$& $^{\,+\,0.0002}_{\,-\,0.0006}$& $^{\,+\,0.002}_{\,-\,0.002}$&  $^{\,+\,0.000}_{\,-\,0.000}$\\ \hline
$\sigma_{\nu_\chi}$& $^{\,+\,0.0017}_{\,-\,0.0016}$ & $^{\,+\,0.0232}_{\,-\,0.0243}$ & $0.034$ & $^{\,+\,0.057}_{\,-\,0.055}$\\ \hline
$\sigma_{\gamma_5}$ & $0.0066$ & $0.0008$ & $0.001$ &  $0.005$\\ \hline
$\sigma_{L_{5,8}}$ & $0.0053$ & $0.0009$ & $0.017$ & $0.015$\\ \hline
$\sigma_{L_{7}}$ & $0.0012$ & $0.0000$ & $0.065$ &  $0.066$\\ \hline
$\sigma_{K_{i}}$& $0.0019$ & $0.0031$ & $0.018$ &  $0.013$\\ \hline
$\sigma_{X_{i}}$ & $0.0020$ & $0.0003$ & $0.003$ & $0.005$\\
\hline
\end{tabular}
\caption{\label{tabnlo}NLO central values for $\alpha\neq 0$ and their parametric errors. }
\end{table}

\section{Exploring dependence on  ``isospin scheme"}
\label{app:scheme}
In this appendix we explore the dependence of $\Omega_{\rm eff}$ on the  
scheme-dependent definition of isospin limit in QCD.
For recent developments on the definition of ``isospin-symmetric QCD"  on the lattice, we refer the reader to Refs.~\cite{Aoki:2019cca,DiCarlo:2019thl}
and references therein.
In our work we  use as reference scheme (``Scheme I") the one adopted in Ref.~\cite{Cirigliano:2003gt}, 
in which the  meson masses in the isospin limit are taken as follows: 
\bea
M_\pi^2  &\equiv & M_{\pi^0}^2~,
\\
M_K^2 & \equiv & M_{K^0}^2~.
\eea
The LO meson masses with inclusion of isospin breaking then read:
\bea
 M_{\pi^0}^2 &=& M_\pi^2~,  
\\
 M_{\pi^\pm}^2 &=& M_\pi^2   + 2 e^2 Z F^2~,
 \\
 M_{K^0}^2 &=&  M_K^2~, 
 \\
  M_{K^\pm}^2 &=&  M_K^2   - \frac{4\, \varepsilon^{(2)}}{\sqrt{3}}    \left( M_K^2 - M_\pi^2 \right)   + 2 e^2 Z F^2~,
\eea
where we  used  $B_0 (m_s  - \hat{m}) = M_K^2  -M_\pi^2  + \mathcal{O}(\varepsilon^{(2)})$ in the second term of $M_{K^\pm}^2$. 
In the hadronic schemes of Refs.~\cite{Aoki:2019cca,DiCarlo:2019thl} this would correspond to defining iso-symmetric  QCD by fixing $\hat{m}$ and $m_s$ from the physical values of $M_{\pi^0}$ and $M_{K^0}$.

We will contrast the above scheme to ``Scheme II", which treats the kaon masses more symmetrically. In this scheme we take the meson masses in the isospin limit to be 
as follows: 
\bea
M_\pi^2  &\equiv & M_{\pi^0}^2~,
\\
M_K^2 & \equiv & \frac{1}{2}  \left\lbrace M_{K^\pm}^2  + M_{K^0}^2    -  ( M_{\pi^\pm}^2 - M_{\pi^0}^2 )   \right\rbrace~.
\label{eq:isoscheme2}
\eea
The LO meson masses with isospin breaking are then 
\bea
 M_{\pi^0}^2 &=& M_\pi^2~,  
\\
 M_{\pi^\pm}^2 &=& M_\pi^2   + 2 e^2 Z F^2~,
 \\
 M_{K^0}^2 &=&  M_K^2  + \frac{2 \,\varepsilon^{(2)}}{\sqrt{3}}  \left( M_K^2 - M_\pi^2 \right)~,
 \\
  M_{K^\pm}^2 &=&  M_K^2   - \frac{2\, \varepsilon^{(2)}}{\sqrt{3}}    \left( M_K^2 - M_\pi^2 \right)    + 2 e^2 Z F^2~,
\eea
where again we used  $B_0 (m_s  - \hat{m}) = M_K^2  -M_\pi^2 + \mathcal{O}(\varepsilon^{(2)})$  to re-write the terms proportional to $\varepsilon^{(2)}$. 
In the hadronic schemes of Refs.~\cite{Aoki:2019cca,DiCarlo:2019thl},  
this would correspond to defining iso-symmetric  QCD by fixing $\hat{m}$ and $m_s$ from the physical values of $M_{\pi^0}$ and  the combination $M_K$ defined by Eq.~(\ref{eq:isoscheme2}).
Note that in Scheme II, to LO in the chiral expansion,  $\hat{m}$ and $m_s$ take the same value in both full QCD and iso-symmetric QCD. This is not the case in Scheme I.

\subsection{Leading-order analysis}

After putting the external legs on the appropriate mass-shells,    the tree-level amplitudes  are:
\bea
A_{+-} &=& - \,\sqrt{2}\, G_8\, F  \left( M_{K^0}^2  - M_{\pi^\pm}^2  - e^2 F^2   g_{\rm ewk}  \right)~, 
\\
A_{00} &=& - \,\sqrt{2}\, G_8\, F  \left( M_{K^0}^2  - M_{\pi^0}^2  \right)   \left(1 - \frac{2}{\sqrt{3}} \varepsilon^{(2)} \right)~,
\\
A_{+0}  &=&  - \,G_8\, F  \left( M_{\pi^0}^2 - M_{\pi^\pm}^2  - e^2 F g_{\rm ewk} \right) 
\nonumber \\
& &  - G_8\, F  \left\lbrace M_{K^\pm}^2 - M_{\pi^0}^2      + \frac{1}{2}  \Big(  
M_{\pi^\pm}^2 - M_{\pi^0}^2    \Big)  \right\rbrace   \,  \frac{2}{\sqrt{3}}\, \varepsilon^{(2)} ~, 
\eea
where the explicit terms involving $\varepsilon^{(2)}$ arise from $\pi^0$-$\eta$ mixing.  
Using the two schemes defined above for the mesons masses, we can 
 split the amplitudes as follows
 \beq
 A_{ij} = A_{ij}^{(0)}  + \delta A_{ij}~, 
 \eeq
where $A_{ij}^{(0)}$ represents the ``isospin limit" result and $\delta A_{ij}$ the deviation from that limit. 
Both terms in the above decomposition are scheme dependent. 

The isospin-limit amplitudes  have the same form in  both schemes:
\beq
A_{+-}^{(0)} = A_{00}^{(0)} = - \,\sqrt{2}\, G_8\,  F \,(M_K^2 - M_\pi^2)~, 
\qquad  \qquad A_{+0}^{(0)} = 0~.
\eeq
The scheme dependence is due to the fact that $M_K^2$  takes different values in the two schemes. 

Using Scheme I, the deviations from the isospin limit are:
\bea
\delta A_{+-}^{({\text{I}})}   &=&   \sqrt{2} \, G_8\, F\, (e^2 F^2)\, ( 2 Z + g_{\rm ewk}) ~,
\\
\delta A_{00}^{{(\text{I}})} &=&   \sqrt{2} \, G_8\, F\, (M_K^2 - M_\pi^2)  \, \frac{2}{\sqrt{3}}\,\varepsilon^{(2)}~,
\\
\delta A_{+0}^{({\text{I}})} &=&    G_8 F (e^2 F^2) ( 2 Z + g_{\rm ewk})  - G_8 F (M_K^2 - M_\pi^2)  \, \frac{2}{\sqrt{3}}\,\varepsilon^{(2)}~.
\eea

Using Scheme II we find: 
\bea
\delta A_{+-}^{({\text{II}})}   &=&  \delta A_{+-}^{({\text{I}})}     -   \sqrt{2} \, G_8\, F\, (M_K^2 - M_\pi^2)  \, \frac{2 }{\sqrt{3}}\,\varepsilon^{(2)}~,
\\
\delta A_{00}^{({\text{II}})} &=&\delta A_{00}^{({\text{I}})}    -  \sqrt{2}\,  G_8 \,F \,(M_K^2 - M_\pi^2)  \, \frac{2}{\sqrt{3}}\,\varepsilon^{(2)} =0~,
\\
\delta A_{+0}^{({\text{II}})} &=&  \delta A_{+0}^{({\text{I}})}~.
\eea

For the isospin-basis amplitudes of interest in $\epsilon^\prime$ we then have:
\bea
\delta A_{0}^{({\text{II}})}   &=&  \delta A_{0}^{({\text{I}})}     -   \sqrt{2}\,  G_8\, F\, (M_K^2 - M_\pi^2)  \, \frac{2}{\sqrt{3}}\,\varepsilon^{(2)}~,
\label{eq:da0scheme}
\\
\delta A_{2}^{({\text{II}})} &=&\delta A_{2}^{({\text{I}})}~,   
\\
\delta A_{2}^{+({\text{II}})} &=&  \delta A_{2}^{+({\text{I}})}~.
\eea

Let us now discuss the implications of the above scheme dependence. 
First, note that  since $\delta A_{2}^{({\text{II}})} =\delta A_{2}^{({\text{I}})}$, the fit to ${\rm Re} \,g_{27}$, controlled by the $K^\pm \to \pi^\pm \pi^0$ rate, 
is essentially unchanged. 

For the CP-violating sector, we need to study the scheme dependence of  $\Omega_{\rm IB}$, $\Delta_0$,  and $f_{5/2}$, 
that appear as  correction factors in the formula for $\epsilon^\prime$, namely:
\bea
\Omega_{\rm IB} &=&  \frac{  {\rm Re}\, A_0^{(0)}}{  {\rm Re}\, A_2^{(0)}} \cdot  \frac{{\rm Im} \,\delta A_2^{\rm non-emp}}{{\rm Im}\, A_0^{(0)}}~,
\\
\Delta_0 &=&   \frac{  {\rm Im}  \,\delta A_0}{  {\rm Im}\, A_0^{(0)}} -  \frac{  {\rm Re}\,  \delta A_0}{  {\rm Re}\, A_0^{(0)}}~, 
\\
f_{5/2} &=&  \frac{5}{3} \, \frac{{\rm Re}\, A_{5/2}}{{\rm Re}\,A_{3/2}^{(0)}}~.
\eea
The above quantities are of first order in isospin-breaking parameters ($\varepsilon^{(2)}$ and $e^2$). 
Now note that the scheme dependence of the ``isospin-limit" quantities denoted by the superscript ``$(0)$" is itself of first order in isospin 
breaking. Therefore we conclude that, to first order in isospin breaking the scheme dependence of $\Omega_{\rm IB}$, $\Delta_0$,  and $f_{5/2}$ 
is controlled by the scheme dependence of $\delta A_0$, $\delta A_2^{\rm non-emp}$,  and $\cA_{5/2}$.  
From  the amplitude shifts given above, we therefore conclude that to leading order in the chiral expansion 
\bea
\Omega_{\rm IB}^{({\text{II}})}  &=& \Omega_{\rm IB}^{({\text{I}})}~,  
\\
f_{5/2}^{({\text{II}})}  &=& f_{5/2}^{({\text{I}})}  =0~,
\\ 
\Delta_0^{({\text{II}})}  &=& \Delta_0^{({\text{I}})}     + \frac{{\rm Im} (\delta A_0^{({\text{II}})} - \delta A_0^{({\text{I}})} )   }{ {\rm Im} A_0^{(0)} }    
 -  \frac{{\rm Re    (\delta A_0^{({\text{II}})} - \delta A_0^{({\text{I}})} )  }  }{ {\rm Re} A_0^{(0)} }~.
\eea
Using Eq.~(\ref{eq:da0scheme}),  the explicit form of $A_0^{(0)}$ to leading order 
\beq
A_0^{(0)} =  -\, \sqrt{2} \, F\, (M_K^2 - M_\pi^2)  \, \left( \frac{G_{27}}{9} + G_8 \right) , 
\eeq
and the fact that ${\rm Im}\, G_{27} = 0$,   we find 
\beq
\Delta_0^{({\rm II})}  - \Delta_0^{({\rm I})}    =  \frac{2\, \varepsilon^{(2)}}{\sqrt{3}} \left( 1 - \frac{1}{1 + \frac{1}{9}  \frac{{\rm Re} \,g_{27}}{{\rm Re}\, g_8}}  \right)
\simeq \ \frac{2 \,\varepsilon^{(2)}}{\sqrt{3}} \times   \frac{1}{9}  \frac{{\rm Re}\, g_{27}}{{\rm Re} \,g_8} \sim  8.5 \times  10^{-5}~.
\eeq
The ``isospin-scheme" dependence is comparable to the LO central value induced by strong isospin breaking using Scheme I,   namely  
$\Delta_{0}^{({\rm I})} \big\vert_{{\rm LO},\, \alpha=0} = - 4 \times 10^{-5}$~\cite{Cirigliano:2003gt}. 
Including EM effects one has    $\Delta_{0}^{({\rm I})} \big\vert_{\rm LO} = (8.7 \pm 3.0) \times 10^{-2}$, implying that the scheme dependence in $\Delta_0$ and therefore in $\Omega_{\rm eff}$
(see Eq.~(\ref{eq:omegaeff})) is completely negligible compared to other uncertainties. 

\subsection{Beyond leading order}

As for the LO analysis,  we focus on  the comparison of ``Scheme I" and ``Scheme II" only. 
We note that to first order in isospin breaking and any order in the chiral expansion  
the only amplitudes that can possibly differ between   Scheme I and  Scheme II are  $\cA_{1/2}^{(\varepsilon)}$ and 
$\cA_{3/2}^{(\varepsilon)}$. 
Based on this observation we already conclude that 
\beq
f_{5/2}^{({\rm II})} = f_{5/2}^{({\rm I})} 
\eeq
holds beyond leading order. 
In order to quantify the isospin-scheme dependence of $\cA_{1/2,3/2}^{(\varepsilon)}$ at NLO, 
we need to consider three effects:
\begin{enumerate}
\item  Expressing $F$ in terms of $F_\pi$ in the tree-level amplitudes;
\item  Counterterm amplitudes proportional to $G_8 N_i$;
\item Loop amplitudes  with $G_8$ insertions and isospin breaking only in the masses (internal and external).  
\end{enumerate}
In what follows we discuss the first two effects. 
For this discussion, let us recall the relevant terms in Eq.~(\ref{eq:generalamplitude})
\beq
\cA_n  \supset  \mbox{} -  G_8 \,F_\pi\,  (M_K^2 - M_\pi^2 ) \left[
\cA_n^{(8)}   + \varepsilon^{(2)} \cA_n^{(\varepsilon)} 
\right] , \qquad \qquad  n=1/2,3/2~.
\eeq

\subsubsection[Expressing $F$ in terms of $F_\pi$ in the tree-level amplitudes]{\boldmath Expressing $F$ in terms of $F_\pi$ in the tree-level amplitudes}

The relation between $F$ and $F_\pi$ takes the form
\beq
\label{eq:Fpi}
F = F_\pi  \, \Big\{ 1 + f^{(s)}  (M_K^2,M_\pi^2)   + \varepsilon^{(2)}  \, g^{(s)} (M_K^2,  M_\pi^2) \Big\}~, \qquad \qquad s= {\rm I, II}~,
\eeq
where  $f^{(s)}(x,y)$ and $g^{(s)}(x,y)$ are scheme-dependent  functions of the meson masses arising from loops and counterterms, 
and $M_K^2$ and $M_\pi^2$ denote the isospin-limit masses in the chosen scheme.  Using the expression of $F_\pi$ in terms of the quark masses~\cite{Gasser:1984gg}, one obtains 
\bea
f^{({\rm I})} (x,y) &=& f^{({\rm II})} (x,y)\, =\, f(x,y)~,\\
g^{({\rm I})} (x,y) &= &  g(x,y)\, =\, 
\frac{2}{\sqrt{3}} \, (x - y)  \left[ \frac{8 L_4^r (\mu)}{F^2}   - \frac{1}{2 (4 \pi F)^2}  \left(1 + \log \frac{x}{\mu^2} \right) \right]~, 
\\
g^{({\rm II})} (x,y) &= & 0~,
\eea
and the form of $f(x,y)$ is irrelevant for our discussion 

Upon making the  substitutions (\ref{eq:Fpi})  in the tree-level amplitudes, one obtains
\beq
\cA_n^{(\varepsilon)}\, =\,   \bar{\cA}_n^{(\varepsilon)}  \,\Big[ 1 +  f^{(s)}(M_K^2,M_\pi^2)  \Big]  \    +   \  a_n^{(8)}   \, g^{(s)} (M_K^2, M_\pi^2) ~ , 
\eeq
where $  \bar{\cA}_n^{(\varepsilon)} $ is the  strong isospin-violating amplitude before making the replacement $F \to F_\pi$. 
The term involving $f(M_K^2,M_\pi^2)$ is scheme independent to first order in isospin breaking (recall that  $\cA_n^{(\varepsilon)}$ is already 
multiplied by $\varepsilon^{(2)}$, so changing the value of the masses in the argument of $f(x,y)$  leads to higher-order effects in isospin breaking). 
The term proportional to $g(x,y)$ is scheme dependent. 
So one gets
\bea
\cA_n^{(\varepsilon),({\rm I})} - \cA_n^{(\varepsilon),({\rm II})}   &=& a_n^{(8)}  \,  g(M_K^2,M_\pi^2) ~.
\eea
Recalling that   
\beq
a_{1/2}^{(8)} = \sqrt{2}~,  
\qquad\qquad
a_{3/2}^{(8)} = 0~, 
\eeq
then one sees that there is no scheme dependence in the $\Delta I= 3/2$ amplitudes, while there is a residual scheme dependence 
in the $\Delta I = 1/2$ amplitude, namely:
\bea
\delta A_2^{(\varepsilon),({\rm II})} & = &\delta A_2^{(\varepsilon),({\rm I})}~,
\\
\delta A_0^{(\varepsilon),({\rm II})} & = &\delta A_0^{(\varepsilon),({\rm I})} + \varepsilon^{(2)}  \, \sqrt{2} \,G_8\, F_\pi\,  (M_K^2 - M_\pi^2)  \, g(M_K^2, M_\pi^2) ~.
\eea
The above results lead to:
\bea
\Omega_{\rm IB}^{({\rm II})}  &=& \Omega_{\rm IB}^{({\rm I})}~,   
\\
\Delta_0^{({\rm II})} -  \Delta_0^{({\rm I})}    &=& 
  \varepsilon^{(2)}  \ g(M_K^2,M_\pi^2) \   \left( 1 - \frac{1}{1 + \frac{1}{9}  \frac{{\rm Re}\, g_{27}}{{\rm Re}\, g_8}}  \right)
  \nonumber \\
&\simeq & \ 
 \varepsilon^{(2)}  \, g(M_K^2, M_\pi^2) \,   \frac{1}{9}  \frac{{\rm Re}\, g_{27}}{{\rm Re}\, g_8} \sim 10^{-6}.
\eea
This is to be compared  to  the  NLO results~\cite{Cirigliano:2003gt} 
 $\Delta_{0}^{({\rm I})} \big\vert_{{\rm NLO},\alpha=0} =  -(5.1 \pm 1.2 ) \times 10^{-3}$
and 
 $\Delta_{0}^{({\rm I})} \big\vert_{\rm NLO}  =  (5.7 \pm 1.7) \times 10^{-2}$, 
 showing again that the scheme dependence 
 of $\Delta_0$ and, therefore,   $\Omega_{\rm eff}$
(see Eq.~(\ref{eq:omegaeff}))    is well below current uncertainties in $\Delta_0$ and $\Omega_{\rm eff}$.

\subsubsection[Contributions proportional to $G_8 N_i$]{\boldmath Contributions proportional to $G_8\, N_i$}

These amplitudes have the structure:
\beq
\cA  \propto  \sum_{i=5}^9   \, N_i \left( \sum_q   A_{i q}   B_0  m_q \right)   \left( \sum_{ab}   B_{ab}  \ p_a \cdot p_b \right)   + 
\sum_{i=10}^{13}   \, N_i \left( \sum_q   {C}_{i q}   B_0 m_q \right)    \left( \sum_q   {D}_{i q}  B_0 m_q \right) , 
\eeq
where $p_n$ are the external particle momenta. 
The  ``isospin scheme" dependence arises when expressing  $p_i \cdot  p_j$ and   $B_0\, m_q$ in terms of the meson masses.  

Expanding the amplitudes in the two schemes  one can check that $\delta A_{+-}$ and $\delta A_{00}$ are shifted by the same amount, 
so only $\delta A_0$ can depend on the scheme. Explicitly we find
\bea
\delta A_2^{(\varepsilon),({\rm II})} & = &\delta A_2^{(\varepsilon),({\rm I})}~,
\\
\delta A_0^{(\varepsilon),({\rm II})} & = &\delta A_0^{(\varepsilon),({\rm I})} -    \ \frac{4\, \varepsilon^{(2)}}{\sqrt{3}}   \ \sqrt{2}\,  F_\pi\,  (M_K^2 - M_\pi^2)    \   \tilde{\Delta} ~,
\\ 
\tilde{\Delta} &=&  \frac{1}{F_\pi^2}  \,G_8  \Big[  M_K^2    \left( 2\, N_5  - 4\,  N_7 + 4\, N_8 + 2\, N_9 \right)  
\nonumber \\
& & \hskip 1.cm\mbox{}
+  \ M_\pi^2   \left( N_5 + 6 N_7  - N_8 - N_9  - 2\, N_{10}  - 4\, N_{11}  - 2\, N_{12} \right) 
\Big]~. 
\eea

As before, the implications for $\epsilon^\prime$ are that $\Omega_{{\rm IB}}$ is scheme independent (up to second order in isospin breaking) 
while $\Delta_0$ is scheme dependent.  Using the above expressions, the scheme dependence of $\Delta_0$ can be estimated as follows: 
\beq
\Delta_0^{({\rm II})} -  \Delta_0^{({\rm I})}    =  \varepsilon^{(2)} \left[ \frac{{\rm Im}  ( \tilde{\Delta} )  }{{\rm Im}\, G_8}  - 
 \frac{{\rm Re} ( \tilde{\Delta} ) }{   {\rm Re}\, G_8  \left(   1 + \frac{1}{9}\,\frac{{\rm Re}\,g_{27}}{{\rm Re} \,g_8}   \right)} 
 \right] \simeq  10^{-3}~, 
 \label{eq:schemeN}
\eeq
still well below the total uncertainty of $\Delta_0$ and $\Omega_{\rm eff}$.

\newpage

\bibliographystyle{JHEP}
\bibliography{mibib}

\end{document}